\documentclass[11pt]{article}
\usepackage{amsmath,amssymb}
\usepackage{amsthm}
\usepackage{ascmac}
\usepackage{float}
\usepackage[dvipdfmx]{graphicx}
\usepackage{comment}
\usepackage[all]{xy}
\usepackage[bookmarksnumbered=true,colorlinks=true,filecolor=blue,linkcolor=blue,urlcolor=blue,citecolor=red,linktocpage=true]{hyperref}

\allowdisplaybreaks

\numberwithin{equation}{section}

\newcommand{\be}{\begin{equation}}
\newcommand{\ee}{\end{equation}}
\newcommand{\ba}{\begin{aligned}}
\newcommand{\ea}{\end{aligned}}

\newcommand{\bs}{\begin{subequations}}
\newcommand{\es}{\end{subequations}}

\begin{document}


\renewcommand{\thefootnote}{\fnsymbol{footnote}}
\setcounter{page}{0}
\thispagestyle{empty}
\thispagestyle{empty}
\begin{flushright}
OCU-PHYS 497
\\
USTC-ICTS-19-05
\end{flushright} 

\vskip3cm
\begin{center}
{\LARGE {\bf Topological String Geometry}}  
\vskip1.5cm
{\large 
\bf {Matsuo Sato\footnote{\href{mailto:msato@hirosaki-u.ac.jp}{msato@hirosaki-u.ac.jp}}}
and
\bf {Yuji Sugimoto\footnote{\href{mailto:sugimoto@ustc.edu.cn}{sugimoto@ustc.edu.cn}}
} 

\vskip1cm

{\it $~^*$Graduate School of Science and Technology, Hirosaki University\\ 
 Bunkyo-cho 3, Hirosaki, Aomori 036-8561, Japan}
 }

\vskip.5cm

 \it $~^\dagger$Osaka City University Advanced Mathematical Institute (OCAMI)
3-3-138, Sugimoto, Sumiyoshi-ku, Osaka, 558-8585, Japan 
\\
 \it $~^\dagger$ Interdisciplinary Center for Theoretical Study, University of Science and Technology of China, Hefei, Anhui 230026, China
\end{center}

\vskip1cm
\begin{abstract} 

Perturbative string amplitudes are correctly derived from the string geometry theory, which is one of the candidates of a non-perturbative formulation of string theory. In order to derive non-perturbative effects rather easily, we formulate topological string geometry theory.  We derive the perturbative partition function of the topological string theory from fluctuations around a classical solution in the topological string geometry theory.

\end{abstract}

\renewcommand{\thefootnote}{\arabic{footnote}}
\setcounter{footnote}{0}

\vfill\eject

\tableofcontents

\section{Introduction}

String geometry theory is one of  the candidates of non-perturbative formulation of string theory \cite{Sato:2017qhj}.  Actually, the theory possesses appropriate properties as a non-perturbative formulation as follows. First, we can derive the all-order perturbative scattering amplitudes that possess the super moduli in IIA, IIB and SO(32) I superstring theories from the single theory by considering fluctuations around fixed perturbative IIA, IIB and SO(32) I vacua, respectively.  Second, the theory is background independent. Third, the theory unifies particles and the space-time. 

Next task is to derive non-perturbative effects from the theory. In order to  derive non-perturbative effects rather easily, we formulate a topological string geometry theory in this paper. It is worthy to study the topological string geometry theory, since non-perturbative partition functions of the topological string theory on a certain class of the Calabi--Yau manifolds are proposed as we will explain below.

Non-perturbative partition functions in topological string theory on non-compact toric Calabi--Yau manifolds were conjectured by using dualities \cite{Lockhart:2012vp, Hatsuda:2013oxa}. In \cite{Hatsuda:2013oxa},  a non-perturbative free energy is given by the combination of the unrefined free energy and Nekrasov--Shatashvili limit of the refined free energy  \cite{Nekrasov:2009rc}. The authors in  \cite{Grassi:2014zfa} proposed that the non-perturbative free energy in \cite{Hatsuda:2013oxa} closely relates to the quantization of the mirror curve \cite{Aganagic:2003qj, Aganagic:2011mi}, based on the duality between ABJM matrix model and the topological string theory on local $\mathbb{P}^1\times\mathbb{P}^1$ \cite{Aganagic:2002wv, Aharony:2008ug, Marino:2009jd}. This relation is sometimes called as Topological String/Spectral Theory correspondence. (The overview of the correspondence is given in e.g. \cite{Marino:2015nla}).  In addition, the authors in \cite{Couso-Santamaria:2016vwq} show that the non-perturbative free energy in \cite{Hatsuda:2013oxa} agrees with that obtained by applying the resurgence technique, developed in \cite{Santamaria:2013rua, Couso-Santamaria:2014iia}, to the perturbative topological string theory in case of local $\mathbb{P}^2$.  In spite of such progresses, the first principle calculations of the non-perturbative partition functions are still not known. We expect to provide an answer to this issue from topological string geometry theory.

The rest of the paper is organized as follows.  In section \ref{TopStrGeom}, we perform a topological twist of string geometry theory and  define topological string geometry theory. In section \ref{Perturbative}, we derive perturbative topological string theory in the flat background by considering fluctuations around a classical solution of the topological string geometry theory. In Appendix \ref{TS}, we develop a superfield formalism of the topological string since the string geometry theory is defined in a superfield formalism.




\section{Topological string geometry}\label{TopStrGeom}
In this section, we will define topological string geometry by performing a topological twist of string geometry \cite{Sato:2017qhj}.  

As in \cite{Sato:2017qhj}, on the topological super Riemannian surfaces $\bar{\bold{\Sigma}}$, there exists an unique Abelian differential $dp$ that has simple poles with residues $f^i$ at $P^i$ where $\sum_i f^i=0$, if it is normalized to have purely imaginary periods with respect to all contours to fix ambiguity of adding holomorphic differentials. A global time is defined by $\bar{z}=\bar{\tau}+i\bar{\sigma}:=\int^{P} dp$ at any point $P$ on $\bar{\bold{\Sigma}}$ \cite{Krichever:1987a, Krichever:1987b}. $\bar{\tau}$ takes the same value at the same point even if different contours are chosen in $\int^P dp$, because the real parts of the periods are zero by definition of the normalization. In particular, $\bar{\tau}=-\infty$ at $P^i$ with negative $f^i$ and $\bar{\tau}=\infty$ at $P^i$ with positive $f^i$. A contour integral on $\bar{\tau}$ constant line around $P^i$: $i \Delta \bar{\sigma}=\oint dp=2\pi i f^i$ indicates that the $\bar{\sigma}$ region around $P^i$ is $2\pi f^i$. This means that $\bar{\bold{\Sigma}}$ around $P^i$ represents a semi-infinite supercylinder with radius $f^i$. The condition $\sum_i f^i=0$ means that the total $\bar{\sigma}$ region of incoming supercylinders equals to that of outgoing ones if we choose the outgoing direction as positive. That is, the total $\bar{\sigma}$ region is conserved. In order to define the above global time uniquely, we fix the $\bar{\sigma}$ regions $2\pi f^i$ around $P^i$. We divide $N$ $P^i$'s to arbitrary two sets consist of $N_-$ and $N+$ $P^i$'s, respectively ($N_{-}+N_{+}=N$), then we divide equally $-1$ to $f^i = \frac{-1}{N_-}$, and $1$ to $f^i = \frac{1}{N_+}$.

Thus, under a superconformal transformation, one obtains a topological super Riemann surface $\bar{\bold{\Sigma}}$ that has coordinates composed of the global time $\bar{\tau}$ and the position $\bar{\sigma}$. Because $\bar{\bold{\Sigma}}$  can be a supermoduli of  super Riemann surfaces \cite{Dijkgraaf:1990qw}, any two-dimensional topological super Riemannian manifold $\bold{\Sigma}$ can be obtained by $\bold{\Sigma}=\psi(\bar{\bold{\Sigma}})$ where $\psi$ is a superdiffeomorphism times super Weyl transformation.

Next, we will define a model space $E$.  We consider a state $(\bar{\bold{\Sigma}}, \Phi_T(\bar{\tau}_s), \bar{\Phi}_T(\bar{\tau}_s) , \bar{\tau}_s)$ determined by $\bar{\bold{\Sigma}}$, a $\bar{\tau}=\bar{\tau}_s$ constant hypersurface and an arbitrary map $(\Phi_T(\bar{\tau}_s),\bar{\Phi}_T(\bar{\tau}_s))$ from $\bar{\bold{\Sigma}}|_{\bar{\tau}_s}$ to the Euclidean space $\bold{R}^d$. $\Phi_T(\bar{\tau})$ and $\bar{\Phi}_T(\bar{\tau})$ are defined by restricting (\ref{Phi}) to the $\bar{\tau}$ constant hypersurface whose conditions are given by
\begin{eqnarray}
&&\bar{h}^{\bar{\tau}m}\partial_{m} \phi +2 i \chi^{z^*}_z \rho_{z^*}=0, \nonumber \\
&&\bar{h}^{\bar{\tau}m}\partial_{m} \bar{\phi} +2 i \chi_{z^*}^z \rho_z=0,
\label{restrictionA}
\end{eqnarray}
for $T=A$ and
\begin{eqnarray}
&&\bar{h}^{\bar{\tau}m}\partial_{m} \phi=0, \nonumber \\
&&\bar{h}^{\bar{\tau}m}\partial_{m} \bar{\phi}  +2 i \chi^{z^*}_z \rho_{z^*}+2 i \chi_{z^*}^z \rho_z=0,
\label{restrictionB}
\end{eqnarray}
for $T=B$ where $\bar{h}_{mn}$ is the metric of the worldsheet, and  $\chi^{z^*}_z$ and $\chi^{z}_{z^*}$ is the gravitino of the worldsheet.

\begin{figure}[htb]
\centering
 \includegraphics[width=12cm]{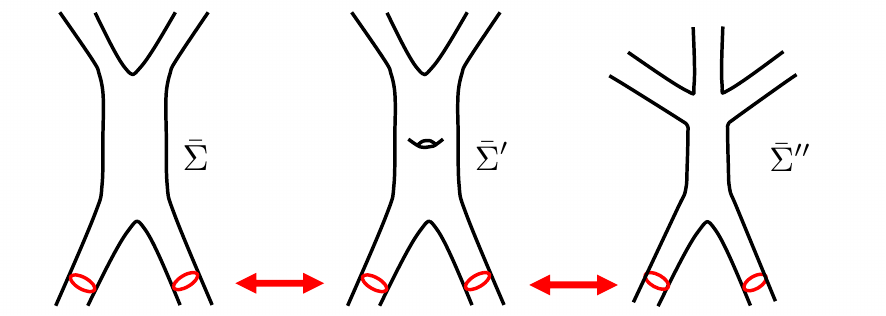}
\caption{An equivalence class of a string state. If the supercylinders and the embedding functions are the same at $\bar{\tau} \simeq -\infty$, the states of strings at $\bar{\tau} \simeq -\infty$ specified by the red lines $(\bar{\bold{\Sigma}}, \Phi_T(\bar{\tau}_s\simeq -\infty), \bar{\Phi}_T(\bar{\tau}_s\simeq -\infty), \bar{\tau}_s\simeq -\infty)$, $(\bar{\bold{\Sigma}}', \Phi_T'(\bar{\tau}_s\simeq -\infty), \bar{\Phi}_T'(\bar{\tau}_s\simeq -\infty), \bar{\tau}_s\simeq -\infty)$, and $(\bar{\bold{\Sigma}}'', \Phi_T''(\bar{\tau}_s\simeq -\infty), \bar{\Phi}_T''(\bar{\tau}_s\simeq -\infty), \bar{\tau}_s\simeq -\infty)$ should be identified.}
     \label{EquivalentClass}
\end{figure}

\begin{figure}[htb]
\centering
\includegraphics[width=3cm]{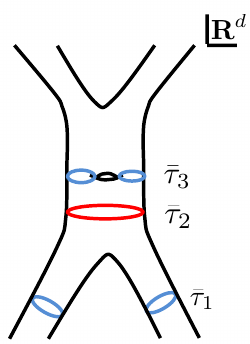}
\caption{Various string states. The red and blue lines represent one-string and two-string states, respectively.}
\label{states}
\end{figure}

$\bar{\bold{\Sigma}}$ is a union of $N_{\pm}$ supercylinders with radii $f_i$ at $\bar{\tau}\simeq \pm \infty$. Thus, we define a string state as an equivalence class $[\bar{\bold{\Sigma}}, \Phi_T(\bar{\tau}_s\simeq \pm \infty),\bar{\Phi}_T(\bar{\tau}_s\simeq \pm \infty), \bar{\tau}_s\simeq \pm \infty]$ by a relation $(\bar{\bold{\Sigma}}, \Phi_T(\bar{\tau}_s\simeq \pm \infty), \bar{\Phi}_T(\bar{\tau}_s\simeq \pm \infty), \bar{\tau}_s\simeq \pm \infty) \sim (\bar{\bold{\Sigma}}', \Phi_T'(\bar{\tau}_s\simeq \pm \infty), \bar{\Phi}_T'(\bar{\tau}_s\simeq \pm \infty), \bar{\tau}_s\simeq \pm \infty)$ if $N_{\pm}=N'_{\pm}$, $f_i=f'_i$, $\Phi_T(\bar{\tau}_s\simeq \pm \infty)=\Phi_T'(\bar{\tau}_s\simeq \pm \infty)$, and $\bar{\Phi}_T(\bar{\tau}_s\simeq \pm \infty)=\bar{\Phi}_T'(\bar{\tau}_s\simeq \pm \infty)$ as in Fig. \ref{EquivalentClass}. Because $\bar{\Sigma}|_{\bar{\tau}_s} \simeq S^1 \cup S^1 \cup \cdots \cup S^1$ where $\bar{\Sigma}$ is the reduced space of $\bar{\bold{\Sigma}}$,  and $(\Phi_T(\bar{\tau}_s),\bar{\Phi}_T(\bar{\tau}_s)): \bar{\bold{\Sigma}}|_{\bar{\tau}_s} \to M$, $[\bar{\bold{\Sigma}}, \Phi_T(\bar{\tau}_s),\bar{\Phi}_T(\bar{\tau}_s), \bar{\tau}_s]$ represent many-body states of strings in $\bold{R}^d$ as in Fig. \ref{states}.
 The model space $E$ is defined by a collection of all the string states as $E:=\cup_T\{[\bar{\bold{\Sigma}}, \Phi_T(\bar{\tau}_s),\bar{\Phi}_T(\bar{\tau}_s), \bar{\tau}_s]\}$, where $T$ runs A and B.


Here, we will define topologies of $E$. We define an $\epsilon$-open neighbourhood of $[\bar{\bold{\Sigma}}, \Phi_{Ts}(\bar{\tau}_s), \bar{\Phi}_{Ts}(\bar{\tau}_s), \bar{\tau}_s]$ by
\be
\ba
&U([\bar{\bold{\Sigma}}, \Phi_{Ts}(\bar{\tau}_s), \bar{\Phi}_{Ts}(\bar{\tau}_s), \bar{\tau}_s], \epsilon)
\\
&\qquad:=
\left\{[\bar{\bold{\Sigma}},  \Phi_T(\bar{\tau}), \bar{\Phi}_T(\bar{\tau}), \bar{\tau}] \bigm| \sqrt{|\bar{\tau}-\bar{\tau}_s|^2+  (\Phi_T(\bar{\tau})- \Phi_{Ts}(\bar{\tau}_s))\cdot (\bar{\Phi}_T(\bar{\tau})- \bar{\Phi}_{Ts}(\bar{\tau}_s))} <\epsilon   \right\},
\label{neighbour}
\ea
\ee
where
\begin{eqnarray}
&&(\Phi_T(\bar{\tau})- \Phi_{Ts}(\bar{\tau}_s))\cdot (\bar{\Phi}_T(\bar{\tau})- \bar{\Phi}_{Ts}(\bar{\tau}_s)) \nonumber \\&:=&\int_0^{2\pi}  d\bar{\sigma}\Biggl( 
\Bigl((\phi(\bar{\tau}, \bar{\sigma})-\phi_s(\bar{\tau}_s, \bar{\sigma})) (\bar{\phi}(\bar{\tau}, \bar{\sigma})-\bar{\phi}_s(\bar{\tau}_s, \bar{\sigma}))  \nonumber \\
&+&(\rho_{z^*}(\bar{\tau}, \bar{\sigma})-\rho_{z^* s}(\bar{\tau}_s, \bar{\sigma}))
(\bar{\rho}_z(\bar{\tau}, \bar{\sigma})-\bar{\rho}_{z s}(\bar{\tau}_s, \bar{\sigma})) \nonumber \\
&+&(\chi(\bar{\tau}, \bar{\sigma})-\chi_s(\bar{\tau}_s, \bar{\sigma}))
(\bar{\chi}(\bar{\tau}, \bar{\sigma})-\bar{\chi}_{s}(\bar{\tau}_s, \bar{\sigma})) \nonumber \\
&+&(F(\bar{\tau}, \bar{\sigma})-F_s(\bar{\tau}_s, \bar{\sigma}))
(\bar{F}(\bar{\tau}, \bar{\sigma})-\bar{F}_{s}(\bar{\tau}_s, \bar{\sigma})) \Bigr) \nonumber \\
&+&(\chi_z^{z^*}(\bar{\tau}, \bar{\sigma})-\chi_{z s}^{z^*}(\bar{\tau}_s, \bar{\sigma})) 
(\chi^z_{z^*}(\bar{\tau}, \bar{\sigma})-\chi^z_{z^* s}(\bar{\tau}_s, \bar{\sigma})) \Biggr).
\end{eqnarray}

$U([\bar{\bold{\Sigma}}, \Phi_T(\bar{\tau}_s\simeq \pm \infty), \bar{\Phi}_T(\bar{\tau}_s\simeq \pm \infty), \bar{\tau}_s\simeq \pm \infty], \epsilon)=
U([\bar{\bold{\Sigma}}', \Phi_T'(\bar{\tau}_s\simeq \pm \infty), \bar{\Phi}_T'(\bar{\tau}_s\simeq \pm \infty), \bar{\tau}_s\simeq \pm \infty], \epsilon)$ consistently if $N_{\pm}=N'_{\pm}$, $f_i=f'_i$, $\Phi_T(\bar{\tau}_s\simeq \pm \infty)=\Phi_T'(\bar{\tau}_s\simeq \pm \infty)$, $\bar{\Phi}_T(\bar{\tau}_s\simeq \pm \infty)=\Phi_T'(\bar{\tau}_s\simeq \pm \infty)$, and $\epsilon$ is small enough, because the $\bar{\tau}_s\simeq \pm \infty$ constant hypersurface traverses only supercylinders overlapped by $\bar{\bold{\Sigma}}$ and $\bar{\bold{\Sigma}}'$.

$U$ is defined to be an open set of $E$ if there exists $\epsilon$ such that $U([\bar{\bold{\Sigma}}, \Phi_T(\bar{\tau}_s), \bar{\Phi}_T(\bar{\tau}_s), \bar{\tau}_s], \epsilon) \subset U$ for an arbitrary point $[\bar{\bold{\Sigma}}, \Phi_T(\bar{\tau}_s), \bar{\Phi}_T(\bar{\tau}_s), \bar{\tau}_s] \in U$. 
The topology of $E$ satisfies the axiom of topology. The proof is the same as in \cite{Sato:2017qhj}.

 Although the model space is defined by using the coordinates $[\bar{\bold{\Sigma}},  \Phi_T(\bar{\tau}_s), \bar{\Phi}_T(\bar{\tau}_s), \bar{\tau}_s]$, the model space does not depend on the coordinates, because the model space is a topological space.

By definition of the $\epsilon$-open neighbourhood, arbitrary two string states on a connected  topological super Riemann surface in $\bold{R}^d$ are connected continuously. Thus, there is an one-to-one correspondence between a topological super Riemann surface with punctures in $\bold{R}^d$ and a curve  parametrized by $\bar{\tau}$ from $\bar{\tau}=-\infty$ to $\bar{\tau}=\infty$ on $E$. That is, curves that represent asymptotic processes on $E$ reproduce the right moduli space of the topological super Riemann surfaces in $\bold{R}^d$.

By a general curve parametrized by $t$ on $E$, string states on different topological super Riemann surfaces that have even different numbers of genera, can be connected continuously, as in Fig. 3, whereas different topological super Riemann surfaces that have different numbers of genera cannot be connected continuously in the moduli space of the topological super Riemann surfaces.

\begin{figure}[htb]
\begin{center}
\includegraphics[width=18cm]{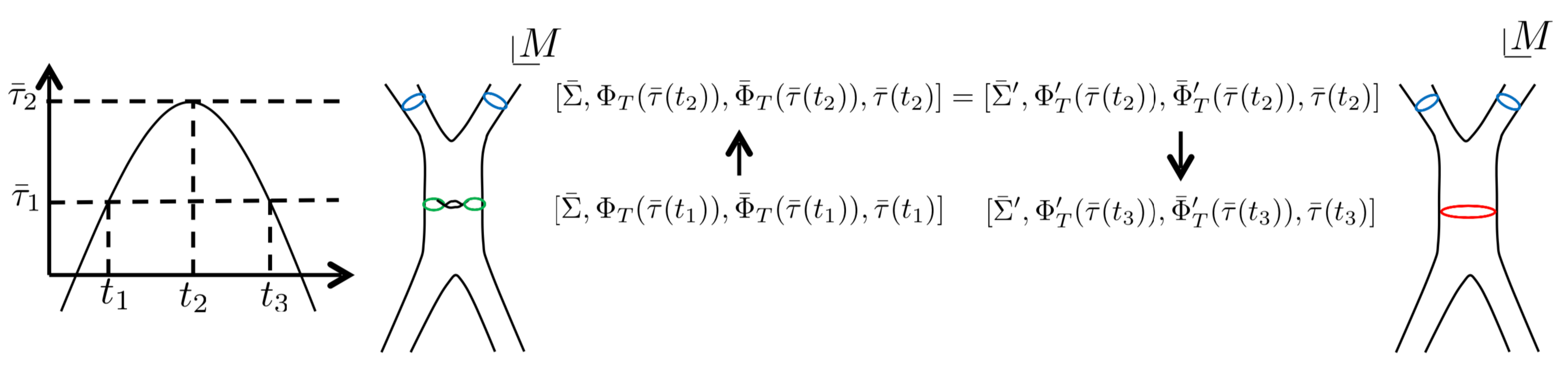}
\caption{A continuous trajectory. 
In case of general $\bar{\tau}(t)$ as in the left graph, string states on different  Riemann surfaces can be connected continuously in $E$ as $[\bar{\bold{\Sigma}}, \Phi_T(\bar{\tau}(t_1)), \bar{\Phi}_T(\bar{\tau}(t_1)), \bar{\tau}(t_1)]$ and $[\bar{\bold{\Sigma}}',\Phi_T'(\bar{\tau}(t_3)),\bar{\Phi}_T'(\bar{\tau}(t_3)), \bar{\tau}(t_3)]$ on the pictures.}
\end{center}
\label{Connected}
\end{figure}

In the following, we denote $[\bar{\bold{E}}_{M}^{\quad A}(\bar{\sigma}, \bar{\tau}, \bar{\theta}^{\alpha}),  \Phi_T(\bar{\tau}),\bar{\Phi}_T(\bar{\tau}), \bar{\tau}]$, where  $\bar{\bold{E}}_{M}^{\quad A}(\bar{\sigma}, \bar{\tau}, \bar{\theta}^{\alpha})$ ($M=(m, \alpha)$, $A=(q, a)$, $m, q=0,1$, $\alpha, a=1,2,3,4$) is the worldsheet topological super vierbein on $\bar{{\boldsymbol \Sigma}}$ defined by (\ref{Phi}),  instead of $[\bar{\bold{\Sigma}},  \Phi_T(\bar{\tau}), \bar{\Phi}_T(\bar{\tau}), \bar{\tau}]$, because giving a topological super Riemann surface is equivalent to giving a topological super vierbein up to super diffeomorphism and super Weyl transformations.

Next, in order to define structures of string manifold, we consider how generally we can define general coordinate transformations between $[\bar{\bold{E}}_{M}^{\quad A}, \Phi_T(\bar{\tau}), \bar{\Phi}_T(\bar{\tau}), \bar{\tau}]$ and $[\bar{\bold{E}}_{M}^{'\quad A}, \Phi_T'(\bar{\tau}'), \bar{\Phi}_T'(\bar{\tau}'), \bar{\tau}']$ where $[\bar{\bold{E}}_{M}^{\quad A}, \Phi_T(\bar{\tau}), \bar{\Phi}_T(\bar{\tau}), \bar{\tau}] \in U \subset E$ and $[\bar{\bold{E}}_{M}^{'\quad A}, \Phi_T'(\bar{\tau}'), \bar{\Phi}_T'(\bar{\tau}'), \bar{\tau}'] \in U'\subset E$. $\bar{\bold{E}}_{M}^{\quad A}$ does not transform to $\bar{\tau}$, $\Phi_T(\bar{\tau})$, and $\bar{\Phi}_T(\bar{\tau})$ and vice versa, because $\bar{\tau}$, $\Phi_T(\bar{\tau})$, and $\bar{\Phi}_T(\bar{\tau})$ are continuous variables, whereas $\bar{\bold{E}}_{M}^{\quad A}$ is a discrete variable: $\bar{\tau}$, $\Phi_T(\bar{\tau})$, and $\bar{\Phi}_T(\bar{\tau})$ vary continuously, whereas $\bar{\bold{E}}_{M}^{\quad A}$ varies discretely in a trajectory on $E$ by definition of the neighbourhoods. $\bar{\tau}$ does not transform to $\bar{\sigma}$ and $\bar{\theta}$ and vice versa, because the string states are defined by $\bar{\tau}$ constant hypersurfaces. Under these restrictions, the most general coordinate transformation is given by 
\be
\ba
&[\bar{\bold{E}}_{M}^{\quad A}(\bar{\sigma}, \bar{\tau}, \bar{\theta}^{\alpha}), \Phi_T(\bar{\sigma}, \bar{\tau},\bar{\theta}),  \bar{\Phi}_T(\bar{\sigma}, \bar{\tau},\bar{\theta}), \bar{\tau}]
\\  
&\mapsto 
[\bar{\bold{E}}_{M}^{'\quad A}(\bar{\sigma}'(\bar{\sigma}, \bar{\theta}), \bar{\tau}'(\bar{\tau}, \Phi_T(\bar{\tau}), \bar{\Phi}_T(\bar{\tau})), \bar{\theta}^{'\alpha}(\bar{\sigma}, \bar{\theta}))), \Phi_T'(\bar{\sigma}', \bar{\tau}', \bar{\theta}')(\bar{\tau}, \Phi_T(\bar{\tau}),\bar{\Phi}_T(\bar{\tau})),
\\
&\qquad\qquad\qquad\qquad
\bar{\Phi}_T'(\bar{\sigma}', \bar{\tau}', \bar{\theta}')(\bar{\tau}, \Phi_T(\bar{\tau}),\bar{\Phi}_T(\bar{\tau})), \bar{\tau}'(\bar{\tau}, \Phi_T(\bar{\tau}), \bar{\Phi}_T(\bar{\tau}))],
 \\
\label{GeneralCoordTrans}
\ea
\ee
where $\bar{\bold{E}}_{M}^{\quad A} \mapsto \bar{\bold{E}}_{M}^{'\quad A}$ represents a world-sheet superdiffeomorphism transformation\footnote{
We extend the model space from $\bold{E}=\{[\bar{\bold{E}}_{M}^{\quad A}(\bar{\sigma}, \bar{\tau}, \bar{\theta}^{\alpha}), \Phi_T(\bar{\sigma}, \bar{\tau},\bar{\theta}),  \bar{\Phi}_T(\bar{\sigma}, \bar{\tau},\bar{\theta}), \bar{\tau}] \}$ to $\bold{E}=\{[\bar{\bold{E}}_{M}^{' \quad A}(\bar{\sigma}', \bar{\tau}', \bar{\theta}^{' \alpha}), \Phi_T'(\bar{\sigma}', \bar{\tau}',\bar{\theta}'),  \bar{\Phi}_T'(\bar{\sigma}', \bar{\tau}',\bar{\theta}'), \bar{\tau}'] \}$ by including the points generated by the superdiffeomorphisms $\bar{\sigma} \mapsto \bar{\sigma}'(\bar{\sigma}, \bar{\theta})$,  $\bar{\theta}^{\alpha} \mapsto \bar{\theta}^{'\alpha}(\bar{\sigma}, \bar{\theta})$, and $\bar{\tau} \mapsto \bar{\tau}'(\bar{\tau})$.}.  $(\bar{\sigma}, \bar{\theta}^{\alpha}) \mapsto (\bar{\sigma}'(\bar{\sigma}, \bar{\theta}), \bar{\theta}^{'\alpha}(\bar{\sigma}, \bar{\theta}))$ represents that these coordinates are transformed by diffeomorphism and only its Q-partners: $\bar{\theta}^{' z^*}=\bar{\theta}^{z^*}$ and $\bar{\theta}^{' z}=\bar{\theta}^{z}$ are not transformed, and $\bar{\sigma}'=\bar{\sigma}'(\bar{\sigma}, \bar{\theta},  \bar{\bar{\theta}})$, $\bar{\theta}'=\bar{\theta}'(\bar{\sigma}, \bar{\theta},  \bar{\bar{\theta}})$ and $\bar{\bar{\theta}}'=\bar{\bar{\theta}}'(\bar{\sigma}, \bar{\theta},  \bar{\bar{\theta}})$ does not depend on $\bar{\theta}^{z^*}$ or $\bar{\theta}^{z}$\footnote{We consider only these diffeomorphism and its Q-partners in the following.}. 
$\Phi_T'(\bar{\tau}, \Phi_T(\bar{\tau}), \bar{\Phi}_T(\bar{\tau}))$, $\bar{\Phi}_T'(\bar{\tau}, \Phi_T(\bar{\tau}), \bar{\Phi}_T(\bar{\tau}))$, and $\bar{\tau}'(\bar{\tau}, \Phi_T(\bar{\tau}), \bar{\Phi}_T(\bar{\tau}))$ are functionals of $\bar{\tau}$, $\Phi_T(\bar{\tau})$, and $\bar{\Phi}_T(\bar{\tau})$. 
We consider all the manifolds which are constructed by patching open sets of the model space $E$ by the general coordinate transformations (\ref{GeneralCoordTrans}) and call them topological string manifolds $\mathcal{M}_D$. An example of the string manifold in the critical string theory is given in \cite{Sato:2017qhj}.

The tangent space is spanned by $\frac{\partial}{\partial\bar{\tau}}$, $\frac{\partial}{\partial \Phi_T^{I}(\bar{\sigma}, \bar{\tau},\bar{\theta})}$, and $\frac{\partial}{\partial \bar{\Phi}_T^{\bar{J}}(\bar{\sigma}, \bar{\tau},\bar{\theta})}$ as one can see from the $\epsilon$-open neighbourhood \eqref{neighbour}\footnote{$\bar{\theta}$ in $\Phi_T^\bold{I}$ denotes the collection of the Grassmann coordinates, $\theta^{z^*}$, $\theta^z$, $\theta$ and $\bar{\theta}$.}.
We should note that $\frac{\partial}{\partial \bar{\bold{E}}^{\quad A}_{M}}$  cannot be a part of basis that span the tangent space because $\bar{\bold{E}}^{\quad A}_{M}$ is just a discrete variable in $\bold{E}$. 
The index of $\frac{\partial}{\partial \Phi_T^{I}(\bar{\sigma}, \bar{\tau},\bar{\theta})}$ and $\frac{\partial}{\partial \bar{\Phi}_T^{\bar{J}}(\bar{\sigma}, \bar{\tau},\bar{\theta})}$ can be $(I \, \bar{\sigma} \bar{\theta})$ and $(\bar{J} \, \bar{\sigma} \bar{\theta})$. 
Then, let us define a summation over $\bar{\sigma}$ and $\bar{\theta}$ that is invariant under $(\bar{\sigma}, \bar{\theta}^{\alpha}) \mapsto (\bar{\sigma}'(\bar{\sigma}, \bar{\theta}), \bar{\theta}^{'\alpha}(\bar{\sigma}, \bar{\theta}))$ and transformed as a scalar under $\bar{\tau} \mapsto \bar{\tau}'(\bar{\tau}, \Phi_T(\bar{\tau}), \bar{\Phi}_T(\bar{\tau}))$.  First, $\int d\bar{\tau} \int d\bar{\sigma}d^4\bar{\theta} \bar{\bold{E}}(\bar{\sigma}, \bar{\tau}, \bar{\theta}^{\alpha})$ is invariant under $(\bar{\sigma}, \bar{\tau}, \bar{\theta}^{\alpha}) \mapsto (\bar{\sigma}'(\bar{\sigma}, \bar{\theta}), \bar{\tau}'(\bar{\tau}, \Phi_T(\bar{\tau}), \bar{\Phi}_T(\bar{\tau})), \bar{\theta}^{'\alpha}(\bar{\sigma}, \bar{\theta}))$, where $\bar{\bold{E}}(\bar{\sigma}, \bar{\tau}, \bar{\theta}^{\alpha})$ is the superdeterminant of $\bar{\bold{E}}_{M}^{\quad A}(\bar{\sigma}, \bar{\tau}, \bar{\theta}^{\alpha})$. 
A super analogue of the lapse function, $\frac{1}{\sqrt{\bar{\bold{E}}^0_A \bar{\bold{E}}^{0 A}}}$ transforms as an one-dimensional vector in the $\bar{\tau}$ direction: 
$\int d\bar{\tau} \frac{1}{\sqrt{\bar{\bold{E}}^0_A \bar{\bold{E}}^{0 A}}}$ is invariant under $\bar{\tau} \mapsto \bar{\tau}'(\bar{\tau}, \Phi_T(\bar{\tau}), \bar{\Phi}_T(\bar{\tau}))$ and transformed as a superscalar under $(\bar{\sigma}, \bar{\theta}^{\alpha}) \mapsto (\bar{\sigma}'(\bar{\sigma}, \bar{\theta}), \bar{\theta}^{'\alpha}(\bar{\sigma}, \bar{\theta}))$. Therefore, 
$\int d\bar{\sigma}d^4\bar{\theta} \hat{\bold{E}}(\bar{\sigma}, \bar{\tau}, \bar{\theta}^{\alpha})$,
where
$\hat{\bold{E}}(\bar{\sigma}, \bar{\tau}, \bar{\theta}^{\alpha})
:=
\sqrt{\bar{\bold{E}}^0_A \bar{\bold{E}}^{0 A}}\bar{\bold{E}}(\bar{\sigma}, \bar{\tau}, \bar{\theta}^{\alpha})$,
is transformed as a scalar under $\bar{\tau} \mapsto \bar{\tau}'(\bar{\tau}, \Phi_T(\bar{\tau}), \bar{\Phi}_T(\bar{\tau}))$ and invariant under $(\bar{\sigma}, \bar{\theta}^{\alpha}) \mapsto (\bar{\sigma}'(\bar{\sigma}, \bar{\theta}), \bar{\theta}^{'\alpha}(\bar{\sigma}, \bar{\theta}))$.

Riemannian topological string manifold is obtained by defining a metric, which is a section of an inner product on the tangent space. The general form of a metric is given by
\begin{align}
ds^2 &(\bar{\bold{E}}, \Phi_T(\bar{\tau}),\bar{\Phi}_T(\bar{\tau}), \bar{\tau}) \nonumber \\
=&G(\bar{\bold{E}}, \Phi_T(\bar{\tau}), \bar{\Phi}_T(\bar{\tau}), \bar{\tau})_{dd} (d\bar{\tau})^2 
\nonumber \\
&+d\bar{\tau} \int d\bar{\sigma} d^4\bar{\theta}  \hat{\bold{E}}  \sum_{I}  G(\bar{\bold{E}}, \Phi_T(\bar{\tau}), \bar{\Phi}_T(\bar{\tau}), \bar{\tau})_{d \; (I \bar{\sigma} \bar{\theta})} d \Phi_T^{I}(\bar{\sigma}, \bar{\tau},\bar{\theta}) \nonumber \\
&+d\bar{\tau} \int d\bar{\sigma} d^4 \bar{\theta}  \hat{\bold{E}}  \sum_{\bar{J}}  G(\bar{\bold{E}}, \Phi_T(\bar{\tau}), \bar{\Phi}_T(\bar{\tau}), \bar{\tau})_{d \; (\bar{J} \bar{\sigma} \bar{\theta})} d \bar{\Phi}_T^{\bar{J}}(\bar{\sigma}, \bar{\tau},\bar{\theta}) \nonumber \\
&+\int d\bar{\sigma} d^4\bar{\theta}   \hat{\bold{E}} \int d\bar{\sigma}' d^4\bar{\theta}' \hat{\bold{E}}' \sum_{I, I'} G(\bar{\bold{E}}, \Phi_T(\bar{\tau}), \bar{\Phi}_T(\bar{\tau}), \bar{\tau})_{ \; (I \bar{\sigma}\bar{\theta})  \; (I' \bar{\sigma}' \bar{\theta}')} d \Phi_T^{I}(\bar{\sigma}, \bar{\tau},\bar{\theta}) d \Phi_T^{I'}(\bar{\sigma}', \bar{\tau},\bar{\theta}') \nonumber \\
&+2\int d\bar{\sigma} d^4\bar{\theta}   \hat{\bold{E}} \int d\bar{\sigma}' d^4\bar{\theta}' \hat{\bold{E}}' \sum_{I, \bar{J}} G(\bar{\bold{E}}, \Phi_T(\bar{\tau}), \bar{\Phi}_T(\bar{\tau}), \bar{\tau})_{ \; (I \bar{\sigma}\bar{\theta})  \; (\bar{J} \bar{\sigma}' \bar{\theta}')} d \Phi_T^{I}(\bar{\sigma}, \bar{\tau},\bar{\theta}) d \bar{\Phi}_T^{\bar{J}}(\bar{\sigma}', \bar{\tau},\bar{\theta}') \nonumber \\
&+\int d\bar{\sigma} d^4\bar{\theta}   \hat{\bold{E}} \int d\bar{\sigma}' d^4\bar{\theta}' \hat{\bold{E}}'  \sum_{\bar{J}, \bar{J}'} G(\bar{\bold{E}}, \Phi_T(\bar{\tau}),\bar{\Phi}_T(\bar{\tau}), \bar{\tau})_{ \; (\bar{J} \bar{\sigma}\bar{\theta})  \; (\bar{J}' \bar{\sigma}' \bar{\theta}')} d \bar{\Phi}_T^{\bar{J}}(\bar{\sigma}, \bar{\tau},\bar{\theta}) d \bar{\Phi}_T^{\bar{J}'}(\bar{\sigma}', \bar{\tau},\bar{\theta}'). \nonumber \\
\end{align}
We summarize the vectors as $d \Phi_T^{\bf I}(\bar{\tau})$ (${\bf I}=d,(I\bar{\sigma}\bar{\theta}),(\bar{J}\bar{\sigma}\bar{\theta})$), where  $d \Phi_T^d(\bar{\tau}):=d\bar{\tau}$, $d \Phi_T^{(I \bar{\sigma}\bar{\theta})}(\bar{\tau}):=d \Phi_T^{I}(\bar{\sigma}, \bar{\tau},\bar{\theta})$, $d \Phi_T^{(\bar{J}\bar{\sigma}\bar{\theta})}(\bar{\tau}):=d \bar{\Phi}_T^{\bar{J}}(\bar{\sigma}, \bar{\tau},\bar{\theta})$. 
Then, the components of the metric are summarized as $G_{{\bf I}{\bf J}}(\bar{\bold{E}}, \Phi_T(\bar{\tau}), \bar{\Phi}_T(\bar{\tau}), \bar{\tau})$. The inverse of the metric $G^{{\bf I}{\bf J}}(\bar{\bold{E}}, \Phi_T(\bar{\tau}), \bar{\Phi}_T(\bar{\tau}), \bar{\tau})$ is defined by $G_{{\bf I}{\bf J}}G^{{\bf J}{\bf K}}=G^{{\bf K}{\bf J}}G_{{\bf J}{\bf I}}=\delta_{\bf I}^{\bf K}$, where $\delta_d^d=1$ and $\delta_{I\bar{\sigma}\bar{\theta}}^{I'\bar{\sigma}' \bar{\theta}'}=\frac{1}{\hat{\bold{E}} } \delta_{I}^{I'}\delta(\bar{\sigma}-\bar{\sigma}')\delta^4(\bar{\theta}-\bar{\theta}'),~\delta_{\bar{J}\bar{\sigma}\bar{\theta}}^{\bar{J}'\bar{\sigma}' \bar{\theta}'}=\frac{1}{\hat{\bold{E}} } \delta_{\bar{J}}^{\bar{J}'}\delta(\bar{\sigma}-\bar{\sigma}')\delta^4(\bar{\theta}-\bar{\theta}'),~\delta_{I\bar{\sigma}\bar{\theta}}^{\bar{J}\bar{\sigma}' \bar{\theta}'}=\delta_{\bar{I}\bar{\sigma}\bar{\theta}}^{J\bar{\sigma}' \bar{\theta}'}=0$. The components of the Riemannian curvature tensor are given by $R^{\bf I}_{{\bf J}{\bf K}{\bf L}}$ in the basis $\frac{\partial}{\partial \Phi_T^{ \bold{I}}(\bar{\tau})}$. The components of the Ricci tensor are $R_{{\bf I}{\bf J}}:=R^{\bf K}_{{\bf I}{\bf K}{\bf J}}=R^d_{{\bf I}d{\bf J}}+\int d\bar{\sigma} d^4\bar{\theta} \hat{\bold{E}} R^{(I \bar{\sigma}\bar{\theta})}_{{\bf I} \; (I \bar{\sigma}\bar{\theta}) \; {\bf J}}+\int d\bar{\sigma} d^4\bar{\theta} \hat{\bold{E}} R^{(\bar{I} \bar{\sigma}\bar{\theta})}_{{\bf I} \; (\bar{I} \bar{\sigma}\bar{\theta}) \; {\bf J}}$. The scalar curvature is 
\begin{eqnarray}
R&:=&G^{{\bf I}{\bf J}} R_{{\bf I}{\bf J}} \nonumber \\
&=&G^{dd}R_{dd}
+2 \int d\bar{\sigma} d^4 \bar{\theta} \hat{\bold{E}} G^{d \; (I \bar{\sigma}\bar{\theta})} R_{d \; (I \bar{\sigma}\bar{\theta})} 
+2 \int d\bar{\sigma} d^4 \bar{\theta} \hat{\bold{E}} G^{d \; (\bar{J} \bar{\sigma}\bar{\theta})} R_{d \; (\bar{J} \bar{\sigma}\bar{\theta})} 
\nonumber \\
&&+\int d\bar{\sigma} d^4 \bar{\theta} \hat{\bold{E}}\int d\bar{\sigma}' d^4 \bar{\theta}' \hat{\bold{E}}' G^{(I \bar{\sigma}\bar{\theta}) \; (I'\bar{\sigma}' \bar{\theta}')}R_{(I\bar{\sigma}\bar{\theta}) (I'\bar{\sigma}' \bar{\theta}')}
+\int d\bar{\sigma} d^4 \bar{\theta} \hat{\bold{E}}\int d\bar{\sigma}' d^4 \bar{\theta}' \hat{\bold{E}}' G^{(I\bar{\sigma}\bar{\theta}) \; (\bar{J}\bar{\sigma}' \bar{\theta}')}R_{(I\bar{\sigma}\bar{\theta})(\bar{J}\bar{\sigma}' \bar{\theta}')} 
\nonumber \\
&&+\int d\bar{\sigma} d^4 \bar{\theta} \hat{\bold{E}}\int d\bar{\sigma}' d^4 \bar{\theta}' \hat{\bold{E}}' G^{(\bar{J} \bar{\sigma}\bar{\theta}) \; (I\bar{\sigma}' \bar{\theta}')}R_{(\bar{J}\bar{\sigma}\bar{\theta}) (I\bar{\sigma}' \bar{\theta}')}
+\int d\bar{\sigma} d^4 \bar{\theta} \hat{\bold{E}}\int d\bar{\sigma}' d^4 \bar{\theta}' \hat{\bold{E}}' G^{(\bar{J}\bar{\sigma}\bar{\theta}) \; (\bar{J}'\bar{\sigma}' \bar{\theta}')}R_{(\bar{J}\bar{\sigma}\bar{\theta})(\bar{J}'\bar{\sigma}' \bar{\theta}')} 
 \; . \nonumber 
\end{eqnarray}
The volume is  $\sqrt{G}$, where $G=det (G_{{\bf I}{\bf J}})$.

By using these geometrical objects, we formulate topological string theory non-perturbatively as\footnote{We should note that the coordinates $\Phi_T(\bar{\tau})$ and $\bar{\Phi}_T(\bar{\tau})$ in a string geometry theory are not just target space coordinates, but embedding functions from the worldsheets to the target space.}\footnote{The fact that this theory is formulated by using world-sheets does not imply that it has only perturbative information. For example, string field theories, which are formulated by using world-sheets, have non-perturbative information concerning the tachyon condensation.}
\begin{equation}
Z=\int \mathcal{D}G \mathcal{D}Ae^{-S},
\end{equation}
where
\begin{equation}
S=\frac{1}{G_N}\int \mathcal{D}\bold{E} \mathcal{D} \Phi_T(\bar{\tau}) \mathcal{D}\bar{\Phi}_T(\bar{\tau}) \mathcal{D} \bar{\tau} 
\sqrt{G} (-R +\frac{1}{4} G_N G^{{\bf I}_1 {\bf I}_2} G^{{\bf J}_1 {\bf J}_2} F_{{\bf I}_1 {\bf J}_1} F_{{\bf I}_2 {\bf J}_2} ). \label{Action}
\end{equation}
As an example of sets of fields on the topological string manifolds, we consider the metric and an $u(1)$ gauge field $A_{\bf I}$ whose field strength is given by $F_{{\bf I}{\bf J}}$. The path integral is canonically defined by summing over metrics and gauge fields on $\mathcal{M}$. By definition, the theory is background independent. $\mathcal{D}\bold{E}$ is the invariant measure of the super vierbeins $\bold{E}_{M}^{\quad A}$ on the two-dimensional topological super Riemannian manifolds $\bold{\Sigma}$. $\bold{E}_{M}^{\quad A}$ and $\bar{\bold{E}}_{M}^{\quad A}$ are related to each others by the super diffeomorphism and super Weyl transformations.

Under 
\begin{equation}
(\bar{\tau}, \Phi_T(\bar{\tau}), \bar{\Phi}_T(\bar{\tau})) \mapsto (\bar{\tau}'(\bar{\tau}, \Phi_T(\bar{\tau}), \bar{\Phi}_T(\bar{\tau})), \Phi_T' (\bar{\tau}')(\bar{\tau}, \Phi_T(\bar{\tau}), \bar{\Phi}_T(\bar{\tau})), \bar{\Phi}_T' (\bar{\tau}')(\bar{\tau}, \Phi_T(\bar{\tau}), \bar{\Phi}_T(\bar{\tau}))),
\label{subdiffeo}
\end{equation} 
$G_{{\bf I}{\bf J}}(\bar{\bold{E}}, \Phi_T(\bar{\tau}), \bar{\Phi}_T(\bar{\tau}), \bar{\tau})$ and $A_{\bf I}(\bar{\bold{E}}, \Phi_T(\bar{\tau}), \bar{\Phi}_T(\bar{\tau}), \bar{\tau})$ are transformed as a symmetric tensor and a vector, respectively and the action is manifestly invariant. 

We define $G_{{\bf I}{\bf J}}(\bar{\bold{E}}, \Phi_T(\bar{\tau}), \bar{\Phi}_T(\bar{\tau}), \bar{\tau})$ and $A_{\bf I}(\bar{\bold{E}}, \Phi_T(\bar{\tau}), \bar{\Phi}_T(\bar{\tau}), \bar{\tau})$ so as to transform as scalars under $\bar{\bold{E}}_{M}^{\quad A}(\bar{\sigma}, \bar{\tau}, \bar{\theta}^{\alpha}) \mapsto
\bar{\bold{E}}_{M}^{'\quad A}(\bar{\sigma}'(\bar{\sigma}, \bar{\theta}), \bar{\tau}, \bar{\theta}^{'\alpha}(\bar{\sigma}, \bar{\theta}))$. Under $(\bar{\sigma}, \bar{\theta})$ superdiffeomorphisms: $(\bar{\sigma}, \bar{\theta}^{\alpha}) \mapsto (\bar{\sigma}'(\bar{\sigma}, \bar{\theta}), \bar{\theta}^{'\alpha}(\bar{\sigma}, \bar{\theta}))$, which are equivalent to 
\begin{eqnarray}
&&[\bar{\bold{E}}_{M}^{\quad A}(\bar{\sigma}, \bar{\tau}, \bar{\theta}^{\alpha}), \Phi_T^{I}(\bar{\sigma}, \bar{\tau}, \bar{\theta}^{\alpha}), \bar{\Phi}_T^{\bar{I}}(\bar{\sigma}, \bar{\tau}, \bar{\theta}^{\alpha}), \bar{\tau}] \nonumber \\
&&\mapsto [\bar{\bold{E}}_{M}^{'\quad A}(\bar{\sigma}'(\bar{\sigma}, \bar{\theta}), \bar{\tau}, \bar{\theta}^{'\alpha}(\bar{\sigma}, \bar{\theta})), \Phi_T^{' I}(\bar{\sigma}'(\bar{\sigma}, \bar{\theta}), \bar{\tau}, \bar{\theta}^{'\alpha}(\bar{\sigma}, \bar{\theta}))(\Phi_T^{I}(\bar{\tau}),  \bar{\Phi}_T^{\bar{I}}(\bar{\tau})),  \nonumber \\
&& \qquad \bar{\Phi}_T^{' \bar{I}}(\bar{\sigma}'(\bar{\sigma}, \bar{\theta}), \bar{\tau}, \bar{\theta}^{'\alpha}(\bar{\sigma}, \bar{\theta}))(\Phi_T^{I}(\bar{\tau}),  \bar{\Phi}_T^{\bar{I}}(\bar{\tau})),  \bar{\tau}] \nonumber \\
&&=[\bar{\bold{E}}_{M}^{'\quad A}(\bar{\sigma}'(\bar{\sigma}, \bar{\theta}), \bar{\tau}, \bar{\theta}^{'\alpha}(\bar{\sigma}, \bar{\theta})), \Phi_T^{I}(\bar{\sigma}, \bar{\tau}, \bar{\theta}^{\alpha}),   \bar{\Phi}_T^{\bar{I}}(\bar{\sigma}, \bar{\tau}, \bar{\theta}^{\alpha}), \bar{\tau}], \label{SuperStringGeometryTrans}
\end{eqnarray}
$G_{d \; (I \bar{\sigma}\bar{\theta})}$ is transformed as a superscalar;
\begin{eqnarray}
&&G'_{d \; (I \bar{\sigma}' \bar{\theta}')}(\bar{\bold{E}}', \Phi_T'(\bar{\tau}),  \bar{\Phi}_T'(\bar{\tau}), \bar{\tau})
 \nonumber \\
&=&
G'_{d \; (I \bar{\sigma}' \bar{\theta}')}(\bar{\bold{E}}, \Phi_T'(\bar{\tau}),  \bar{\Phi}_T'(\bar{\tau}), \bar{\tau})
=
\frac{\partial \Phi_T^{\bold{I}}(\bar{\tau})}{\partial \Phi_T^{'d}(\bar{\tau})}
\frac{\partial \Phi_T^{\bold{J}}(\bar{\tau})}{\partial \Phi_T^{'(I \bar{\sigma}' \bar{\theta}')}(\bar{\tau})}
G_{\bold{I} \bold{J}}(\bar{\bold{E}}, \Phi_T(\bar{\tau}),  \bar{\Phi}_T(\bar{\tau}), \bar{\tau})
\nonumber \\
&=&
\frac{\partial \Phi_T^{\bold{I}}(\bar{\tau})}{\partial \Phi_T^{d}(\bar{\tau})}
\frac{\partial \Phi_T^{\bold{J}}(\bar{\tau})}{\partial \Phi_T^{(I \bar{\sigma} \bar{\theta})}(\bar{\tau})}
G_{\bold{I} \bold{J}}(\bar{\bold{E}}, \Phi_T(\bar{\tau}),  \bar{\Phi}_T(\bar{\tau}), \bar{\tau})
=
G_{d \; (I \bar{\sigma} \bar{\theta})}(\bar{\bold{E}}, \Phi_T(\bar{\tau}),  \bar{\Phi}_T(\bar{\tau}), \bar{\tau}), 
\end{eqnarray}
because (\ref{SuperStringGeometryTrans}) and (\ref{subdiffeo}).
The other fields are also transformed as 
\be
\ba
&G'_{d \; (\bar{J} \bar{\sigma}' \bar{\theta}')}(\bar{\bold{E}}', \Phi_T'(\bar{\tau}),  \bar{\Phi}_T'(\bar{\tau}), \bar{\tau})
=
G_{d \; (\bar{J} \bar{\sigma}\bar{\theta})}(\bar{\bold{E}}, \Phi_T(\bar{\tau}), \bar{\Phi}_T(\bar{\tau}), \bar{\tau}),\\ 
&G'_{dd}(\bar{\bold{E}}', \Phi_T'(\bar{\tau}), \bar{\Phi}_T'(\bar{\tau}), \bar{\tau})=G_{dd}(\bar{\bold{E}}, \Phi_T(\bar{\tau}), \bar{\Phi}_T(\bar{\tau}), \bar{\tau}),
\\ 
&G'_{ \; (I \bar{\sigma}' \bar{\theta}')  \; (J \bar{\rho}' \bar{\theta}')}(\bar{\bold{E}}', \Phi_T'(\bar{\tau}), \bar{\Phi}_T'(\bar{\tau}), \bar{\tau})=G_{ \; (I \bar{\sigma}\bar{\theta})  \; (J \bar{\rho}\bar{\theta})}(\bar{\bold{E}},  \Phi_T(\bar{\tau}), \bar{\Phi}_T(\bar{\tau}), \bar{\tau}),
\\
&G'_{ \; (I \bar{\sigma}' \bar{\theta}')  \; (\bar{J} \bar{\rho}' \bar{\theta}')}(\bar{\bold{E}}', \Phi_T'(\bar{\tau}), \bar{\Phi}_T'(\bar{\tau}), \bar{\tau})=G_{ \; (I\bar{\sigma}\bar{\theta})  \; (\bar{J} \bar{\rho}\bar{\theta})}(\bar{\bold{E}},  \Phi_T(\bar{\tau}), \bar{\Phi}_T(\bar{\tau}), \bar{\tau}),
\\
&G'_{ \; (\bar{I} \bar{\sigma}' \bar{\theta}')  \; (\bar{J} \bar{\rho}' \bar{\theta}')}(\bar{\bold{E}}', \Phi_T'(\bar{\tau}), \bar{\Phi}_T'(\bar{\tau}), \bar{\tau})=G_{ \; (\bar{I} \bar{\sigma}\bar{\theta})  \; (\bar{J} \bar{\rho}\bar{\theta})}(\bar{\bold{E}},  \Phi_T(\bar{\tau}), \bar{\Phi}_T(\bar{\tau}), \bar{\tau}),
\ea
\ee
and
\be
\ba
&A'_d(\bar{\bold{E}}', \Phi_T'(\bar{\tau}), \bar{\Phi}_T'(\bar{\tau}), \bar{\tau})=A_d(\bar{\bold{E}},  \Phi_T(\bar{\tau}), \bar{\Phi}_T(\bar{\tau}), \bar{\tau}),
\\
&A'_{(I \bar{\sigma}' \bar{\theta}')}(\bar{\bold{E}}', \Phi_T'(\bar{\tau}), \bar{\Phi}_T'(\bar{\tau}), \bar{\tau}),
=A_{(I \bar{\sigma} \bar{\theta})}(\bar{\bold{E}},  \Phi_T(\bar{\tau}), \bar{\Phi}_T(\bar{\tau}), \bar{\tau}),
\\
&A'_{(\bar{I} \bar{\sigma}' \bar{\theta}')}(\bar{\bold{E}}', \Phi_T'(\bar{\tau}), \bar{\Phi}_T'(\bar{\tau}), \bar{\tau}),
=A_{(\bar{I} \bar{\sigma} \bar{\theta})}(\bar{\bold{E}},  \Phi_T(\bar{\tau}), \bar{\Phi}_T(\bar{\tau}), \bar{\tau}) .
\ea
\ee
Thus, the action is invariant under the $(\bar{\sigma}, \bar{\theta})$ superdiffeomorphisms, because 
\begin{eqnarray}
\int d\bar{\sigma}' d \bar{\theta} '  \hat{\bold{E}}'(\bar{\sigma}', \bar{\tau}, \bar{\theta}')&=&\int d\bar{\sigma} d \bar{\theta}   \hat{\bold{E}}(\bar{\sigma}, \bar{\tau}, \bar{\theta}) 
\end{eqnarray}
 Therefore, $G_{{\bf I}{\bf J}}(\bar{\bold{E}},  \Phi_T(\bar{\tau}), \bar{\Phi}_T(\bar{\tau}), \bar{\tau})$ and $A_{\bf I}(\bar{\bold{E}},  \Phi_T(\bar{\tau}), \bar{\Phi}_T(\bar{\tau}), \bar{\tau})$ are transformed covariantly and the action (\ref{Action}) is invariant under the diffeomorphisms (\ref{GeneralCoordTrans}), including the $(\bar{\sigma}, \bar{\theta})$ Q-superdiffeomorphisms.

\section{Perturbative topological string from topological string geometry}\label{Perturbative}
In this section, from the topological string geometry theory, we will derive the partition function of the A model for topological strings in all-order string coupling constant.  The partition function of the B model can be derived in the same way.

The background that represents a perturbative vacuum for the A model is given by 
\begin{align}
\bar{ds}^2
&= 2\lambda \bar{\rho}(\bar{h}) N^2(\Phi_A(\bar{\tau}), \bar{\Phi}_A(\bar{\tau})) (d \Phi_A^d)^2
\nonumber \\
&\qquad +\int d\bar{\sigma}  \hat{\bold{E}}d^4 \bar{\theta} \int d\bar{\sigma}' \hat{\bold{E}}' d^4 \bar{\theta}' N^{\frac{2}{2-D}}(\Phi_A(\bar{\tau}), \bar{\Phi}_A(\bar{\tau})) \frac{\bar{e}^2(\bar{\sigma}, \bar{\tau})\hat{\bold{E}}(\bar{\sigma}, \bar{\tau}, \bar{\theta})}{\sqrt{\bar{h}
(\bar{\sigma}, \bar{\tau})}}\eta_{(I \bar{\sigma}\bar{\theta} ) (\bar{J}' \bar{\sigma}' \bar{\theta}')}
d \Phi_A^{(I \bar{\sigma}\bar{\theta})} d \bar{\Phi}_A^{(\bar{J}' \bar{\sigma}' \bar{\theta}')}, \nonumber \\
\bar{A}_d&=i \sqrt{\frac{2-2{\bf D}}{2-{\bf D}}}\frac{\sqrt{2\lambda \bar{\rho}(\bar{h}) }}{\sqrt{G_N}} N(\Phi_A(\bar{\tau}), \bar{\Phi}_A(\bar{\tau})), \quad
\bar{A}_{(I \bar{\sigma}\bar{\theta})}=0,\quad \bar{A}_{(\bar{J} \bar{\sigma}\bar{\theta})}=0,
\label{solution}
\end{align}
where we fix charts by choosing $T=A$ on $E$.
$\bar{\rho}(\bar{h}):=\frac{1}{4 \pi}\int d\bar{\sigma} \bar{e}\bar{R}_{\bar{h}}$, where $\bar{R}_{\bar{h}}$ is the scalar curvature of $\bar{h}_{ mn}$. ${\bf D}$ is a volume of the index $(I \bar{\sigma}\bar{\theta})$ and $(\bar{J} \bar{\sigma}\bar{\theta})$: ${\bf D} :=\int d \bar{\sigma} d^4 \bar{\theta} \hat{\bold{E}}\delta_{(I \bar{\sigma}\bar{\theta})}^{ (I \bar{\sigma}\bar{\theta})}+\int d \bar{\sigma} d^4 \bar{\theta} \hat{\bold{E}}\delta_{(\bar{J} \bar{\sigma}\bar{\theta}) }^{(\bar{J} \bar{\sigma}\bar{\theta})}=d 4 \pi \delta(\bar{\sigma}-\bar{\sigma})\delta^4(\bar{\theta}-\bar{\theta})$. $N(\Phi_A(\bar{\tau}), \bar{\Phi}_A(\bar{\tau}))=\frac{1}{1+\alpha v(\Phi_A(\bar{\tau}), \bar{\Phi}_A(\bar{\tau}))}$, 
where
\begin{equation}
v(\Phi_A(\bar{\tau}), \bar{\Phi}_A(\bar{\tau}))= 2\int d \bar{\sigma} d^4 \bar{\theta}  \bar{e} \frac{\sqrt{\bar{\bold{E}}}}{(\bar{h})^{\frac{1}{4}}}(\Phi_A^I(\bar{\tau})\Phi_A^I(\bar{\tau})+\bar{\Phi}_A^{\bar{J}}(\bar{\tau})\bar{\Phi}_A^{\bar{J}}(\bar{\tau})).
\end{equation} 
Then, $v(\Phi_A(\bar{\tau}), \bar{\Phi}_A(\bar{\tau}))$ satisfies
\be
\ba
&\int d\bar{\tau} d\bar{\sigma}  d^4 \bar{\theta} \frac{\sqrt{\bar{h}}}{\bar{e}^2} \eta^{I\bar{J}} \frac{\partial v}{\partial \Phi_A^I(\bar{\tau})}  \frac{\partial v}{\partial \bar{\Phi}_A^{\bar{J}}(\bar{\tau})} 
\\
&\qquad
=4\int d\bar{\tau} d\bar{\sigma}  d^4 \bar{\theta} \bar{\bold{E}}
\eta_{I\bar{J}}\Phi_A^I(\bar{\tau})  \bar{\Phi}_A^{\bar{J}}(\bar{\tau})
\\
&\qquad
=
\int d\bar{\tau} d\bar{\sigma} \eta_{I\bar{J}} \sqrt{\bar{h}}\bigl(
\frac{1}{\bar{e}^2}\partial_{\bar{\sigma}} \phi^{I} \partial_{\bar{\sigma}} \bar{\phi}^{\bar{J}} 
-2\bar{\rho}_z^{\bar{J}}   \partial_{\bar{\sigma}} \chi^{I}
+2\rho^{I}_{z^*}  \partial_{\bar{\sigma}} \bar{\chi}^{\bar{J}}
+F^{I}\bar{F}^{\bar{J}}
\\&\hspace{40mm}
+ i (e^{\bar{\sigma}}_{z^*}+2 \bar{n}^{\bar{\sigma}} ) \bar{\rho}_z^{\bar{J}}  \chi^z_{z^*}\partial_{\bar{\sigma}} \phi^{I}
+ i (e^{\bar{\sigma}}_{z}+2 \bar{n}^{\bar{\sigma}} )  \rho^{I}_{z^*}  \chi_z^{z^*}\partial_{\bar{\sigma}} \bar{\phi}^{\bar{J}}
-4(+\bar{n}^2) \chi^z_{z^*}\chi_z^{z^*} \rho_{z^*}^{I} \bar{\rho}_z^{\bar{J}}
\bigr),
\label{quad2}
\ea
\ee
where we have used (\ref{restrictionA}), and $e_z^{\bar{\sigma}}$ is the vierbein.

The inverse of the metric is given by 
\be
\ba
&\bar{G}^{dd}=\frac{1}{2\lambda \bar{\rho} }\frac{1}{N^2},  \\
&\bar{G}^{d \; (I\bar{\sigma}\bar{\theta})}=\bar{G}^{d \; (\bar{J} \bar{\sigma}\bar{\theta})}=0,  \\
&\bar{G}^{(I\bar{\sigma}\bar{\theta}) \; (\bar{J}\bar{\sigma}'\bar{\theta}')}= N^{\frac{-2}{2-{\bf D}}} \frac{\sqrt{\bar{h}}}{\bar{e}^2\hat{\bold{E}}}  \eta_{(I \bar{\sigma}\bar{\theta}) (\bar{J} \bar{\sigma}'\bar{\theta}')},
\ea
\ee
because $\int d\bar{\sigma}'' d^4{\bar{\theta}''} \hat{\bold{E}}'' \bar{G}_{(I \bar{\sigma}\bar{\theta}) \; (\bar{J}\bar{\sigma}''\bar{\theta}'')}\bar{G}^{(\bar{J}\bar{\sigma}''\bar{\theta}'') \; (I'\bar{\sigma}'\bar{\theta}')}=\int d\bar{\sigma}'' d^4{\bar{\theta}''} \hat{\bold{E}}'' \eta_{(I\bar{\sigma}\bar{\theta}) \; (\bar{J}\bar{\sigma}''\bar{\theta}'')}\eta^{(\bar{J}\bar{\sigma}''\bar{\theta}'') \; (I'\bar{\sigma}'\bar{\theta}')}
= \delta_{(I\bar{\sigma}\bar{\theta})}^{(I'\bar{\sigma}'\bar{\theta}')}$. 
From the metric, we obtain 
\begin{align}
&\sqrt{\bar{G}}=N^\frac{2}{2-{\bf D}}\sqrt{2\lambda \bar{\rho} \exp(\int d\bar{\sigma} d^4 \bar{\theta} \hat{\bold{E}}
(\delta^{I \bar{\sigma} \bar{\theta}}_{I \bar{\sigma} \bar{\theta}}
+\delta^{\bar{I} \bar{\sigma} \bar{\theta}}_{\bar{I} \bar{\sigma} \bar{\theta}})
 \ln \frac{\bar{e}^2\hat{\bold{E}}}{\sqrt{\bar{h}}})} ,
\nonumber \\
&\bar{R}_{dd}=-2\lambda \bar{\rho} N^{\frac{-2}{2-{\bf D}}} \int d\bar{\sigma} d^4 \bar{\theta} \frac{\sqrt{\bar{h}}}{\bar{e}^2} \eta^{I\bar{J}} \partial_{(I \bar{\sigma}\bar{\theta})}N \partial_{(\bar{J} \bar{\sigma}\bar{\theta})}N,
\nonumber \\
&\bar{R}_{d \; (I \bar{\sigma} \bar{\theta})}=0,~\bar{R}_{d \; (\bar{J} \bar{\sigma} \bar{\theta})}=0,
\nonumber \\
&\bar{R}_{(I \bar{\sigma}\bar{\theta}) \; (I' \bar{\sigma}'\bar{\theta}')}
=\frac{{\bf D}-1}{2-{\bf D}}N^{-2}\partial_{(I \bar{\sigma}\bar{\theta})}N \partial_{(I' \bar{\sigma}'\bar{\theta}')}N,
\nonumber \\
&\bar{R}_{(I\bar{\sigma}\bar{\theta}) \; (\bar{J}\bar{\sigma}'\bar{\theta}')}
=\frac{{\bf D}-1}{2-{\bf D}}N^{-2}\partial_{(I \bar{\sigma}\bar{\theta})}N \partial_{(\bar{J} \bar{\sigma}'\bar{\theta}')}N
\nonumber \\
& \qquad \qquad \qquad +\frac{2}{{\bf D}-2}N^{-2}
\int d\bar{\sigma}'' d^4 \bar{\theta}'' 
\frac{\sqrt{\bar{h}''}}{\bar{e}^{''2}}
\eta^{K\bar{L}} \partial_{(K \bar{\sigma}''\bar{\theta}'')}N \partial_{(\bar{L} \bar{\sigma}''\bar{\theta}'')}N
\frac{\hat{\bold{E}} \bar{e}^2}{\sqrt{\bar{h}}}
\eta_{(I\bar{\sigma}\bar{\theta}) \; (\bar{J}\bar{\sigma}'\bar{\theta}')},
\nonumber \\
&\bar{R}=\frac{{\bf D}-3}{2-{\bf D}} N^{\frac{2{\bf D}-6}{2-{\bf D}}}
\int d\bar{\sigma} d^4 \bar{\theta} 
\frac{\sqrt{\bar{h}}}{\bar{e}^2}
\eta^{I\bar{J}} \partial_{(I \bar{\sigma}\bar{\theta})}N \partial_{(\bar{J} \bar{\sigma}\bar{\theta})}N.
\end{align}
By using these quantities, one can show that the background (\ref{solution}) is a classical solution\footnote{This solution is a generalization of the Majumdar-Papapetrou solution \cite{Majumdar1947, Papapetrou1948} of the Einstein-Maxwell system.} to the equations of motion of (\ref{Action}). We also need to use the fact that $v(\Phi_A(\bar{\tau}), \bar{\Phi}_A(\bar{\tau}))$ is a harmonic function with respect to $\Phi_A(\bar{\tau})$ and $\bar{\Phi}_A(\bar{\tau})$, $\eta^{I\bar{J}}\partial_{(I \bar{\sigma}\bar{\theta})}\partial_{(\bar{J} \bar{\sigma}\bar{\theta})}v=0$. In these calculations, we should note that $\bar{\bold{E}}_{M}^{\quad A}$, $\Phi_A^{I}(\bar{\tau})$, $\bar{\Phi}_A^{\bar{J}}(\bar{\tau})$, and $\bar{\tau}$ are all independent. Because the equations of motion are differential equations with respect to $\Phi_A^{I}(\bar{\tau})$, $\bar{\Phi}_A^{\bar{J}}(\bar{\tau})$ and $\bar{\tau}$, $\bar{\bold{E}}_{M}^{\quad A}$ is a constant in the solution (\ref{solution}) to the differential equations. The dependence of $\bar{\bold{E}}_{M}^{\quad A}$ on the background (\ref{solution}) is uniquely determined  by the consistency of the quantum theory of the fluctuations around the background. Actually, we will find that all the perturbative topological  string amplitudes are derived.


Let us consider fluctuations around the background (\ref{solution}), $G_{\bold{I}\bold{J}}=\bar{G}_{\bold{I}\bold{J}}+\tilde{G}_{\bold{I}\bold{J}}$ and $A_\bold{I}=\bar{A}_\bold{I}+\tilde{A}_\bold{I}$. The action (\ref{Action}) up to the quadratic order is given by,
\be
\ba
S=&\frac{1}{G_N} \int \mathcal{D}\bold{E}  \mathcal{D}\Phi_A \mathcal{D}\bar{\Phi}_A \mathcal{D}\bar{\tau}  \sqrt{\bar{G}} 
\Bigl(-\bar{R}+\frac{1}{4}\bar{F}'_{\bold{I}\bold{J}}\bar{F}'^{\bold{I}\bold{J}} 
 \\
&+\frac{1}{4}\bar{\nabla}_\bold{I} \tilde{G}_{\bold{J}\bold{K}} \bar{\nabla}^\bold{I} \tilde{G}^{\bold{J}\bold{K}}
-\frac{1}{4}\bar{\nabla}_\bold{I} \tilde{G} \bar{\nabla}^\bold{I} \tilde{G}
+\frac{1}{2}\bar{\nabla}^\bold{I} \tilde{G}_{\bold{I}\bold{J}} \bar{\nabla}^\bold{J} \tilde{G}
-\frac{1}{2}\bar{\nabla}^\bold{I} \tilde{G}_{\bold{I}\bold{J}} \bar{\nabla}_\bold{K} \tilde{G}^{\bold{J}\bold{K}}
 \\
&-\frac{1}{4}(-\bar{R}+\frac{1}{4}\bar{F}'_{\bold{K}\bold{L}}\bar{F}'^{\bold{K}\bold{L}})
(\tilde{G}_{\bold{I}\bold{J}}\tilde{G}^{\bold{I}\bold{J}}-\frac{1}{2}\tilde{G}^2)
+(-\frac{1}{2}\bar{R}^{\bold{I}}_{\;\; \bold{J}}+\frac{1}{2}\bar{F}'^{\bold{I}\bold{K}}\bar{F}'_{\bold{J}\bold{K}})
\tilde{G}_{\bold{I}\bold{L}}\tilde{G}^{\bold{J}\bold{L}}
 \\
&+(\frac{1}{2}\bar{R}^{\bold{I}\bold{J}}-\frac{1}{4}\bar{F}'^{\bold{I}\bold{K}}\bar{F}'^{\bold{J}}_{\;\;\;\; \bold{K}})
\tilde{G}_{\bold{I}\bold{J}}\tilde{G}
+(-\frac{1}{2}\bar{R}^{\bold{I}\bold{J}\bold{K}\bold{L}}+\frac{1}{4}\bar{F}'^{\bold{I}\bold{J}}\bar{F}'^{\bold{K}\bold{L}})
\tilde{G}_{\bold{I}\bold{K}}\tilde{G}_{\bold{J}\bold{L}}
 \\
&+\frac{1}{4}G_N \tilde{F}_{\bold{I}\bold{J}} \tilde{F}^{\bold{I}\bold{J}} 
+\sqrt{G_N} 
(\frac{1}{4} \bar{F}^{'\bold{I}\bold{J}} \tilde{F}_{\bold{I}\bold{J}} \tilde{G} 
-\bar{F}^{'\bold{I}\bold{J}} \tilde{F}_{\bold{I}\bold{K}} \tilde{G}_\bold{J}^{\;\; \bold{K}} ) \Bigr), \label{fluctuation}
\ea
\ee
where $\bar{F}'_{\bold{I}\bold{J}}:=\sqrt{G_N}\bar{F}_{\bold{I}\bold{J}}$ is independent of $G_N$. $\tilde{G}:=\bar{G}^{\bold{I}\bold{J}}\tilde{G}_{\bold{I}\bold{J}}$. There is no first order term because the background satisfies the equations of motion. If we take $G_N \to 0$, we obtain 
\be
\ba
S'=&\frac{1}{G_N} \int \mathcal{D} \bold{E}   \mathcal{D}\Phi_A \mathcal{D}\bar{\Phi}_A \mathcal{D}\bar{\tau}  \sqrt{\bar{G}} 
\Bigl(-\bar{R}+\frac{1}{4}\bar{F}'_{\bold{I}\bold{J}}\bar{F}'^{\bold{I}\bold{J}} 
 \\
&+\frac{1}{4}\bar{\nabla}_\bold{I} \tilde{G}_{\bold{J}\bold{K}} \bar{\nabla}^\bold{I} \tilde{G}^{\bold{J}\bold{K}}
-\frac{1}{4}\bar{\nabla}_\bold{I} \tilde{G} \bar{\nabla}^\bold{I} \tilde{G}
+\frac{1}{2}\bar{\nabla}^\bold{I} \tilde{G}_{\bold{I}\bold{J}} \bar{\nabla}^\bold{J} \tilde{G}
-\frac{1}{2}\bar{\nabla}^\bold{I} \tilde{G}_{\bold{I}\bold{J}} \bar{\nabla}_\bold{K} \tilde{G}^{\bold{J}\bold{K}}
 \\
&-\frac{1}{4}(-\bar{R}+\frac{1}{4}\bar{F}'_{\bold{K}\bold{L}}\bar{F}'^{\bold{K}\bold{L}})
(\tilde{G}_{\bold{I}\bold{J}}\tilde{G}^{\bold{I}\bold{J}}-\frac{1}{2}\tilde{G}^2)
+(-\frac{1}{2}\bar{R}^{\bold{I}}_{\;\; \bold{J}}+\frac{1}{2}\bar{F}'^{\bold{I}\bold{K}}\bar{F}'_{\bold{J}\bold{K}})
\tilde{G}_{\bold{I}\bold{L}}\tilde{G}^{\bold{J}\bold{L}}
 \\
&+(\frac{1}{2}\bar{R}^{\bold{I}\bold{J}}-\frac{1}{4}\bar{F}'^{\bold{I}\bold{K}}\bar{F}'^{\bold{J}}_{\;\;\;\; \bold{K}})
\tilde{G}_{\bold{I}\bold{J}}\tilde{G}
+(-\frac{1}{2}\bar{R}^{\bold{I}\bold{J}\bold{K}\bold{L}}+\frac{1}{4}\bar{F}'^{\bold{I}\bold{J}}\bar{F}'^{\bold{K}\bold{L}})
\tilde{G}_{\bold{I}\bold{K}}\tilde{G}_{\bold{J}\bold{L}} \Bigr),
\ea
\ee
where the fluctuation of the gauge field is suppressed. In order to fix the gauge symmetry (\ref{subdiffeo}), we take the harmonic gauge. If we add the gauge fixing term
\begin{equation}
S_{fix}=\frac{1}{G_N}\int \mathcal{D}\bold{E}   \mathcal{D}\Phi_A \mathcal{D}\bar{\Phi}_A  \mathcal{D}\bar{\tau}  \sqrt{\bar{G}} 
\frac{1}{2} \Bigl( \bar{\nabla}^\bold{J}(\tilde{G}_{\bold{I}\bold{J}}-\frac{1}{2}\bar{G}_{\bold{I}\bold{J}}\tilde{G}) \Bigr)^2,
\end{equation}
we obtain
\be
\ba
S'+S_{fix}&=\frac{1}{G_N} \int \mathcal{D}\bold{E} \mathcal{D}\Phi_A \mathcal{D}\bar{\Phi}_A  \mathcal{D}\bar{\tau}  \sqrt{\bar{G}} 
\Bigl(-\bar{R}+\frac{1}{4}\bar{F}'_{\bold{I}\bold{J}}\bar{F}'^{\bold{I}\bold{J}} 
 \\
&+\frac{1}{4}\bar{\nabla}_\bold{I} \tilde{G}_{\bold{J}\bold{K}} \bar{\nabla}^\bold{I} \tilde{G}^{\bold{J}\bold{K}}
-\frac{1}{8}\bar{\nabla}_\bold{I} \tilde{G} \bar{\nabla}^\bold{I} \tilde{G}
 \\
&-\frac{1}{4}(-\bar{R}+\frac{1}{4}\bar{F}'_{\bold{K}\bold{L}}\bar{F}'^{\bold{K}\bold{L}})
(\tilde{G}_{\bold{I}\bold{J}}\tilde{G}^{\bold{I}\bold{J}}-\frac{1}{2}\tilde{G}^2)
+(-\frac{1}{2}\bar{R}^{\bold{I}}_{\;\; \bold{J}}+\frac{1}{2}\bar{F}'^{\bold{I}\bold{K}}\bar{F}'_{\bold{J}\bold{K}})
\tilde{G}_{\bold{I}\bold{L}}\tilde{G}^{\bold{J}\bold{L}}
 \\
&+(\frac{1}{2}\bar{R}^{\bold{I}\bold{J}}-\frac{1}{4}\bar{F}'^{\bold{I}\bold{K}}\bar{F}'^{\bold{J}}_{\;\;\;\; \bold{K}})
\tilde{G}_{\bold{I}\bold{J}}\tilde{G}
+(-\frac{1}{2}\bar{R}^{\bold{I}\bold{J}\bold{K}\bold{L}}+\frac{1}{4}\bar{F}'^{\bold{I}\bold{J}}\bar{F}'^{\bold{K}\bold{L}})
\tilde{G}_{\bold{I}\bold{K}}\tilde{G}_{\bold{J}\bold{L}} \Bigr).
\label{fixedaction}
\ea
\ee

In order to obtain perturbative topological string amplitudes, we perform a derivative expansion of $\tilde{G}_{\bold{I}\bold{J}}$,
\be
\ba
&\tilde{G}_{\bold{I}\bold{J}} \to \frac{1}{\alpha} \tilde{G}_{\bold{I}\bold{J}},  \\
&\partial_{\bold{K}}\tilde{G}_{\bold{I}\bold{J}} \to \partial_{\bold{K}}\tilde{G}_{\bold{I}\bold{J}},  \\
&\partial_{\bold{K}}\partial_{\bold{L}}\tilde{G}_{\bold{I}\bold{J}} \to \alpha \partial_{\bold{K}}\partial_{\bold{L}} \tilde{G}_{\bold{I}\bold{J}},
\ea
\ee
and take
\begin{equation}
\alpha \to 0,
\end{equation}
where $\alpha$ is an arbitrary constant in the solution (\ref{solution}). 

We normalize the fields as $\tilde{H}_{\bold{I}\bold{J}}:=Z_{\bold{I}\bold{J}} \tilde{G}_{\bold{I}\bold{J}}$, where $Z_{\bold{I}\bold{J}}:=\frac{1}{\sqrt{G_N}} 
\bar{G}^{\frac{1}{4}} 
(\bar{a}_\bold{I} \bar{a}_\bold{J})^{-\frac{1}{2}}$. $\bar{a}_{\bold{I}}$ represent the background metric as $\bar{G}_{\bold{I}\bold{J}}=\bar{a}_\bold{I} \delta_{\bold{I}\bold{J}}$, where $\bar{a}_d=2\lambda\bar{\rho}$ and $\bar{a}_{(I \bar{\sigma} \bar{\theta})}=\frac{\bar{e}^2\hat{\bold{E}}}{\sqrt{\bar{h}}}$. Then, (\ref{fixedaction}) reduces to
\begin{equation}
S'+S_{fix} \to S_0 + S_2,
\end{equation}
where
\begin{equation}
S_0
=
\frac{1}{G_N} \int \mathcal{D} \bold{E} \mathcal{D}\mathcal{D}\Phi_A \mathcal{D}\bar{\Phi}_A  \mathcal{D}\bar{\tau}  \sqrt{\bar{G}} 
\Bigl(-\bar{R}+\frac{1}{4}\bar{F}'_{\bold{I}\bold{J}}\bar{F}'^{\bold{I}\bold{J}} \Bigr),
\end{equation}
and
\begin{equation}
S_2
=
\int \mathcal{D} \bold{E} \mathcal{D}\Phi_A \mathcal{D}\bar{\Phi}_A  \mathcal{D}\bar{\tau}
\frac{1}{8}\tilde{H}_{\bold{I}\bold{J}}H_{{\bold{I}\bold{J}};\bold{K}\bold{L}}\tilde{H}_{\bold{K}\bold{L}}.
\end{equation}

In the same way as in \cite{Sato:2017qhj}, a part of the action 
\begin{equation}
\int \mathcal{D} \bold{E} \mathcal{D}\Phi_A \mathcal{D}\bar{\Phi}_A  \mathcal{D}\bar{\tau} \frac{1}{4}
\int_0^{2\pi}d\bar{\sigma}d^4 \bar{\theta} 
\eta^{I\bar{J}}\tilde{H}^{\bot}_{d(I \bar{\sigma} \bar{\theta})} 
H
\tilde{H}^{\bot}_{d(\bar{J} \bar{\sigma} \bar{\theta})} \label{SuperSecondOrderAction}
\end{equation}
with 
\begin{eqnarray}
H
=
-\frac{1}{2}\frac{1}{2\lambda\bar{\rho}}(\frac{\partial}{\partial \bar{\tau}})^2
-\int_0^{2\pi} d \bar{\sigma} \int d^4 \bar{\theta} \frac{\sqrt{\bar{h}}}{\bar{e}^2}  \eta^{I \bar{J}}\frac{\partial}{\partial \Phi_A^{I}} \frac{\partial}{\partial \bar{\Phi}_A^{\bar{J}}}
+\frac{\bold{D}^2-5\bold{D}+8}{(2-\bold{D})^2}
\int_0^{2\pi} d \bar{\sigma} \int d^4 \bar{\theta} \frac{\sqrt{\bar{h}}}{\bar{e}^2} \eta^{I \bar{J}}\frac{\partial v}{\partial \Phi_A^I}  \frac{\partial v}{\partial \bar{\Phi}_A^{\bar{J}}}  
\nonumber \\
\end{eqnarray}
decouples from the other modes.

In the following, we consider a sector that consists of local fluctuations in a sense of strings as 
\be
\ba
\tilde{H}'_{\bold{K}\bold{L}} =  \int d \bar{\sigma} d^4 \bar{\theta}  \hat{\bold{E}}f_{\bold{K}\bold{L}}(\bar{\bold{E}}, \Phi_A(\bar{\tau}), \bar{\Phi}_A(\bar{\tau}), \bar{\tau}).
\ea
\ee
In the same way as in \cite{Sato:2017qhj}, we have 
\begin{equation}
\int d\bar{\sigma} \int d^4\bar{\theta} \hat{\bold{E}}(\frac{1}{\bar{e}}\frac{\partial}{\partial \Phi_A^{I}})  (\frac{1}{\bar{e}}\frac{\partial}{\partial \bar{\Phi}_A^{\bar{J}}}) \tilde{H}'_{\bold{K}\bold{L}}
=
\int d\bar{\sigma} \hat{\bold{E}} (\frac{1}{\bar{e}}\frac{\partial}{\partial \phi^{I}})  (\frac{1}{\bar{e}}\frac{\partial}{\partial \bar{\phi}^{\bar{J}}}) \tilde{H}'_{\bold{K}\bold{L}},
\end{equation}
because the leading term of $\Phi_A^I$ is $\phi^I$ and covariant derivatives with respect to $\bar{\sigma}$ apply to all the other terms including $\phi^I$ in $\Phi_A^I$. The same is true of $\bar{\Phi}_A^{\bar{I}}$.

By adding to (\ref{SuperSecondOrderAction}), 
\be
\ba
0
=
\int \mathcal{D}\bold{E} \mathcal{D}\Phi_A \mathcal{D}\bar{\Phi}_A \mathcal{D}\bar{\tau} \frac{1}{4}
\int_0^{2\pi}d\bar{\sigma}'d^4\bar{\theta}' 
\eta^{I\bar{J}}
\tilde{H}^{\bot}_{d(I \bar{\sigma}' \bar{\theta}')} 
\biggl( \int_0^{2\pi} d \bar{\sigma}
\bar{n}^{\bar{\sigma}}
(\partial_{\bar{\sigma}}  \phi^I  \frac{\partial}{\partial \phi^{I}} + \partial_{\bar{\sigma}}   \bar{\phi}^{\bar{J}}  \frac{\partial}{\partial \bar{\phi}^{\bar{J}}} )\bigg)
\tilde{H}^{\bot}_{d(\bar{J} \bar{\sigma}' \bar{\theta}')}, 
 \\
\ea
\ee
and
\be
\ba
0
=
&\int \mathcal{D}\bold{E} \mathcal{D}\Phi_A \mathcal{D}\bar{\Phi}_A \mathcal{D}\bar{\tau} \frac{1}{2}
\int_0^{2\pi}d\bar{\sigma}'d^4\bar{\theta}' 
\eta^{I\bar{J}}
\tilde{H}^{\bot}_{d(I \bar{\sigma}' \bar{\theta}')} 
 \\
&\qquad
\biggl( \int_0^{2\pi} d \bar{\sigma}
\sqrt{\bar{h}}  \bar{n}\bigl(
i \rho_{z^*}^{K} \chi_z^{z^*}  (-i\frac{1}{\bar{e}}\frac{\partial}{\partial \phi^K })
 +
i \bar{\rho}_z^{\bar{L}} \chi^z_{z^*}(-i\frac{1}{\bar{e}}\frac{\partial}{\partial  \bar{\phi}^{\bar{L}}  })\bigr)\biggr)
\tilde{H}^{\bot}_{d(\bar{J} \bar{\sigma}' \bar{\theta}')}, 
 \\
\ea
\ee
where $\bar{n}$ and $\bar{n}_{\bar{\sigma}}$ are components of $\bar{h}$ in the ADM formalism, for example summarized in \cite{Sato:2017qhj}, 
we obtain (\ref{SuperSecondOrderAction}) 
with
\be
\ba
&H(-i\frac{\partial}{\partial \bar{\tau}}, -i\frac{1}{\bar{e}}\frac{\partial}{\partial \phi},-i\frac{1}{\bar{e}}\frac{\partial}{\partial \bar{\phi}}, \Phi_A(\bar{\tau}), \bar{\Phi}_A(\bar{\tau}), \bar{\bold{E}}) \\
&\qquad=
\frac{1}{2}\frac{1}{2\lambda\bar{\rho}}(-i\frac{\partial}{\partial \bar{\tau}})^2
+\int d\bar{\sigma} \Biggl(\sqrt{\bar{h}}\eta^{I\bar{J}}\biggl((-i\frac{1}{\bar{e}}\frac{\partial}{\partial \phi^{I}})(-i\frac{1}{\bar{e}}\frac{\partial}{\partial \bar{\phi}^{\bar{J}}})+2 i \bar{n}\bigl(
\rho_{z^*}^{I} \chi_z^{z^*}  (-i\frac{1}{\bar{e}}\frac{\partial}{\partial \phi^{I} })
 +
\bar{\rho}_z^{\bar{J}} \chi^z_{z^*}(-i\frac{1}{\bar{e}}\frac{\partial}{\partial  \phi^{\bar{J}} })\bigr)\biggr) 
\\
& \qquad\qquad
+i \bar{e} \bar{n}^{\bar{\sigma}} (-i\frac{1}{\bar{e}})(\partial_{\bar{\sigma}}  \phi^I  \frac{\partial}{\partial \phi^{I}} + \partial_{\bar{\sigma}}   \bar{\phi}^{\bar{J}}  \frac{\partial}{\partial \bar{\phi}^{\bar{J}}} )\Biggr)
+\int d\bar{\sigma} d^4\bar{\theta}  \hat{\bold{E}}\eta^{I\bar{J}} \frac{\partial v}{\partial \Phi_A^I}  \frac{\partial v}{\partial \bar{\Phi}_A^{\bar{J}}},
\label{SuperbosonicHamiltonian}
\ea
\ee
where we have taken $\bold{D} \to \infty$.

The propagator $\Delta_F(\bar{\bold{E}}, \Phi_A(\bar{\tau}), \bar{\Phi}_A(\bar{\tau}), \bar{\tau}; \; \bar{\bold{E}}', \Phi_A'(\bar{\tau}') ,\bar{\Phi}_A'(\bar{\tau}'), \bar{\tau}')$ for $\tilde{H}^{\bot}_{d(\bar{J} \bar{\sigma} \bar{\theta})}$ defined as
\be
\ba
\Delta_F(\bar{\bold{E}}, \Phi_A(\bar{\tau}), \bar{\Phi}_A(\bar{\tau}), \bar{\tau}; \; \bar{\bold{E}}', \Phi_A'(\bar{\tau}') ,\bar{\Phi}_A'(\bar{\tau}'), \bar{\tau}')
=
\langle \tilde{H}^{\bot}_{d(\bar{J} \bar{\sigma} \bar{\theta})}(\bar{\bold{E}},\Phi_A(\bar{\tau}),\bar{\Phi}_A(\bar{\tau}),\bar{\tau}) \tilde{H}^{\bot}_{d(\bar{J} \bar{\sigma} \bar{\theta})}(\bar{\bold{E}}',\Phi_A'(\bar{\tau}'),\bar{\Phi}_A'(\bar{\tau}'),\bar{\tau}') \rangle
\ea
\ee
 satisfies 
\be
\ba
&H(-i\frac{\partial}{\partial \bar{\tau}}, -i\frac{1}{\bar{e}}\frac{\partial}{\partial \phi}, -i\frac{1}{\bar{e}}\frac{\partial}{\partial \bar{\phi}}, \Phi_A(\bar{\tau}), \bar{\Phi}_A(\bar{\tau}), \bar{\bold{E}})
\Delta_F(\bar{\bold{E}}, \Phi_A(\bar{\tau}), \bar{\Phi}_A(\bar{\tau}), \bar{\tau}; \; \bar{\bold{E}}', \Phi_A'(\bar{\tau}') ,\bar{\Phi}_A'(\bar{\tau}'), \bar{\tau}')
\\ &\qquad
=
\delta(\bar{\bold{E}}-\bar{\bold{E}}')\delta(\Phi_A(\bar{\tau})-\Phi_A'(\bar{\tau}'))\delta(\bar{\Phi}_A(\bar{\tau})-\bar{\Phi}_A'(\bar{\tau}'))\delta(\bar{\tau}-\bar{\tau}'). \label{SuperPropagatorDef}
\ea
\ee

In order to obtain a Schwinger representation of the propagator, we use the operator formalism $(\hat{\bar{\bold{E}}}, \hat{\Phi}_A(\bar{\tau}),\hat{\bar{\Phi}}_A(\bar{\tau}), \hat{\bar{\tau}})$ of the first quantization. The eigen state for $(\hat{\bar{\bold{E}}},  \hat{\phi},\hat{\bar{\phi}}, \hat{\bar{\tau}})$ is given by $|\bar{\bold{E}}, \phi, \bar{\phi}, \bar{\tau}>$. The conjugate momentum is written as $(\hat{p}_{\bar{\bold{E}}}, \hat{p}_{\phi},\hat{p}_{\bar{\phi}}, \hat{p}_{\bar{\tau}})$. There is no conjugate momentum for the auxiliary field $F^{I}$ and $\bar{F}^{\bar{I}}$. The conjugate momentum of $\chi^I$ is $\bar{\rho}_I$. The normalized operators $\hat{\tilde{\chi}}^{I}$ and its conjugate momentum $\hat{\tilde{\bar{\rho}}}_I$ satisfy  $\{ \hat{\tilde{\bar{\rho}}}_{I}(\bar{\sigma}), \hat{\tilde{\chi}}^{I'}(\bar{\sigma}') \}= \frac{1}{\bar{e}} \delta^{I'}_I \delta(\bar{\sigma}-\bar{\sigma}')$, $\{ \hat{\tilde{\bar{\rho}}}_{I}(\bar{\sigma}), \hat{\tilde{\bar{\rho}}}_{I'}(\bar{\sigma}') \}= 0,~\{ \hat{\tilde{\chi}}^{I}(\bar{\sigma}), \hat{\tilde{\chi}}^{I'}(\bar{\sigma}') \}= 0$. The vacuum $|0>$ for this algebra is defined by $\hat{\tilde{\chi}}^{I}(\bar{\sigma}) |0>=0$. The eigen state $|\tilde{\chi}>$, which satisfies $\hat{\tilde{\chi}}^{I}(\bar{\sigma}) |\tilde{\chi}>= \tilde{\chi}^{I}(\bar{\sigma}) |\tilde{\chi}>$, is given by 
$e^{-\tilde{\chi} \cdot \hat{\tilde{\bar{\rho}}}} |0>=e^{- \int d\bar{\sigma} \hat{\bold{E}}\tilde{\chi}^{I}(\bar{\sigma}) \hat{\tilde{\bar{\rho}}}_{I}(\bar{\sigma})} |0>$. Then, the inner product is given by $<\tilde{\bar{\rho}} | \tilde{\chi}>=e^{\tilde{\bar{\rho}} \cdot \tilde{\chi}}$, whereas the completeness relation is $\int \mathcal{D}\tilde{\bar{\rho}} \mathcal{D}\tilde{\chi} |\tilde{\bar{\rho}}>e^{-\tilde{\bar{\rho}} \cdot \tilde{\chi}}<\tilde{\chi}|=1$. There are similar equations in $\bar{\chi}^{\bar{I}}$ and its conjugate momentum $\rho_{\bar{I}}$.

Because (\ref{SuperPropagatorDef}) means that $\Delta_F$ is an inverse of $H$, $\Delta_F$ can be expressed by a matrix element of the operator $\hat{H}^{-1}$ as
\begin{eqnarray}
&&\Delta_F(\bar{\bold{E}}, \Phi_A(\bar{\tau}), \bar{\Phi}_A(\bar{\tau}), \bar{\tau}; \; \bar{\bold{E}}', \Phi_A'(\bar{\tau}') ,\bar{\Phi}_A'(\bar{\tau}'), \bar{\tau}') \nonumber \\
&=&
<\bar{\bold{E}}, \Phi_A(\bar{\tau}), \bar{\Phi}_A(\bar{\tau}), \bar{\tau}| \hat{H}^{-1}(\hat{p}_{\bar{\tau}}, \hat{p}_{\phi},\hat{p}_{\bar{\phi}}, \hat{\Phi}_A,\hat{\bar{\Phi}}_A, \hat{\bar{\bold{E}}}) |\bar{\bold{E}}',  \Phi_A'(\bar{\tau}') ,\bar{\Phi}_A'(\bar{\tau}'), \bar{\tau}'>.
\label{SuperInverseH2}
\end{eqnarray}
The formula
\begin{equation}
\hat{H}^{-1} = \int_0 ^\infty d T e^{-T \hat{H}}
\end{equation}
implies that
\begin{eqnarray}
&&\Delta_F(\bar{\bold{E}}, \Phi_A(\bar{\tau}), \bar{\Phi}_A(\bar{\tau}), \bar{\tau}; \; \bar{\bold{E}}', \Phi_A'(\bar{\tau}') ,\bar{\Phi}_A'(\bar{\tau}'), \bar{\tau}') \nonumber \\
&=&
\int _0^{\infty} dT <\bar{\bold{E}},\Phi_A(\bar{\tau}), \bar{\Phi}_A(\bar{\tau}), \bar{\tau}|  e^{-T\hat{H}} |\bar{\bold{E}}',   \Phi_A'(\bar{\tau}') ,\bar{\Phi}_A'(\bar{\tau}'), \bar{\tau}'>.
\end{eqnarray}

In order to define two-point correlation functions that is invariant under the general coordinate transformations in the topological string geometry, we define in and out states as
\be
\ba
||\Phi_{Ai},\bar{\Phi}_{Ai} \,|\, \bold{E}_f, ; \bold{E}_i>_{in}&:= \int_{\bold{E}_i}^{\bold{E}_f} \mathcal{D} \bold{E} '|\bar{\bold{E}},' \Phi_{Ai},\bar{\Phi}_{Ai}, \bar{\tau}=-\infty >, \\
<\Phi_{Af},\bar{\Phi}_{Af}\,|\, \bold{E}_f, ; \bold{E}_i||_{out}&:= \int_{\bold{E}_i}^{\bold{E}_f} \mathcal{D} \bold{E} <\bar{\bold{E}}, \Phi_{Af},\bar{\Phi}_{Af}, \bar{\tau}=\infty|,
\ea
\ee
where $\Phi_{Af}:= \Phi_A(\bar{\tau}=\infty),\bar{\Phi}_{Af} := \bar{\Phi}_A(\bar{\tau}=\infty),\Phi_{Ai} := \Phi_A(\bar{\tau}=-\infty),\bar{\Phi}_{Ai} := \bar{\Phi}_A(\bar{\tau}=-\infty)$, and $\bold{E}_i$ and $\bold{E}_f$ represent the topological super vierbein of the super cylinders at $\bar{\tau}=\mp \infty$, respectively. When we insert asymptotic states, we integrate out $\Phi_{Af},\bar{\Phi}_{Af},\Phi_{Ai}, \bar{\Phi}_{Ai}, \bold{E}_f$, and $\bold{E}_i$ in the two-point correlation function for these states,
\begin{equation}
\Delta_F (\Phi_{Af},\bar{\Phi}_{Af};  \Phi_{Ai},\bar{\Phi}_{Ai} | \bold{E}_f, ; \bold{E}_i) 
:=\int _0^{\infty} dT <\Phi_{Af},\bar{\Phi}_{Af} \,|\, \bold{E}_f, ; \bold{E}_i||_{out}  e^{-T\hat{H}} || \Phi_{Ai},\bar{\Phi}_{Ai} \,|\, \bold{E}_f, ; \bold{E}_i>_{in}.
\end{equation}
In the same way as in \cite{Sato:2017qhj}, by inserting completeness relations of the eigen states, we obtain
\begin{align}
\Delta_F &(\Phi_{Af},\bar{\Phi}_{Af};  \Phi_{Ai},\bar{\Phi}_{Ai} | \bold{E}_f, ; \bold{E}_i) \nonumber \\
&=
\int^{\bold{E}_f, \Phi_{Af},\bar{\Phi}_{Af}, \infty }_{\bold{E}_i, \Phi_{Ai},\bar{\Phi}_{Ai}, -\infty} 
\mathcal{D} T \mathcal{D} \bold{E} \mathcal{D}\bar{\tau} \mathcal{D}\Phi_A \mathcal{D}\bar{\Phi}_A
\int 
\mathcal{D} p_T
\mathcal{D}p_{\bar{\tau}}  \mathcal{D}p_{\phi}  \mathcal{D}p_{\bar{\phi}} \mathcal{D}p_{F} \mathcal{D}p_{\bar{F}} 
\nonumber \\
&e^{- \int_{-\infty}^{\infty} dt (-ip_{T} \cdot \frac{d}{dt} T  -ip_{F} \cdot \frac{d}{dt} F  -ip_{\bar{F}} \cdot \frac{d}{dt} \bar{F} 
+\tilde{\bar{\rho}}\cdot  \frac{d}{dt}\tilde{\chi} +\tilde{\rho}\cdot  \frac{d}{dt}\tilde{\bar{\chi}} - ip_{\bar{\tau}} \frac{d}{dt} \bar{\tau}-ip_{\phi} \cdot \frac{d}{dt} \phi -ip_{\bar{\phi}} \cdot \frac{d}{dt} \bar{\phi} +TH(p_{\bar{\tau}},  p_{\phi}, p_{\bar{\phi}}, \Phi_A(\bar{\tau}), \bar{\Phi}_A(\bar{\tau}), \bar{\bold{E}}))}.
\end{align}
If we integrate out $p_{\bar{\tau}}$, $p_{\phi}$, $p_{\bar{\phi}}$, $p_{F}$, and $p_{\bar{F}}$ by using the relation of the ADM formalism, we obtain
\be
\ba
\Delta_F &(\Phi_{Af},\bar{\Phi}_{Af};  \Phi_{Ai},\bar{\Phi}_{Ai} | \bold{E}_f, ; \bold{E}_i)  \\
&=
\int^{\bold{E}_f, \Phi_{Af},\bar{\Phi}_{Af}, \infty }_{\bold{E}_i, \Phi_{Ai},\bar{\Phi}_{Ai}, -\infty} 
\mathcal{D} T
\mathcal{D} \bold{E} \mathcal{D}\bar{\tau} \mathcal{D}\Phi_A \mathcal{D}\bar{\Phi}_A
 \int \mathcal{D} p_T
 \\
&\qquad \times \exp \Biggl(- \int_{-\infty}^{\infty} dt \Bigl(-i p_{T}(t) \frac{d}{dt} T(t)   +\lambda\bar{\rho}\frac{1}{T(t)}(\frac{d \bar{\tau}(t)}{dt})^2
 +\int d\bar{\sigma} d^4 \bar{\theta}\bar{\bold{E}} T(t) \Phi_A' \bar{\Phi}_A' \Biggr).
\label{intermidiate}
\ea
\ee
When the last equality is obtained, we use (\ref{quad2}) and (\ref{top_actionA}).
In the last line, $F^I$ and $\bar{F}^{\bar{I}}$ are constant with respect to $t$, and $\Phi_A'$ and $\bar{\Phi}_A'$ are given by replacing $\frac{\partial}{\partial \bar{\tau}}$ with $\frac{1}{T(t)}\frac{\partial}{\partial t}$ in $\Phi_A$ and $\bar{\Phi}_A$, respectively. 
The path integral is defined over all possible trajectories with fixed boundary values in $E$.

By inserting $\int \mathcal{D}c \mathcal{D}b e^{\int_0^{1} dt \left(\frac{d b(t)}{dt} \frac{d c(t)}{dt}\right)},$ where $b(t)$ and $c(t)$ are bc ghosts, we obtain 
\be
\ba
&\Delta_F (\Phi_{Af},\bar{\Phi}_{Af};  \Phi_{Ai},\bar{\Phi}_{Ai} | \bold{E}_f, ; \bold{E}_i) 
 \\
&~~=
\bold{Z}_0\int^{\bold{E}_f, \Phi_{Af},\bar{\Phi}_{Af}, \infty }_{\bold{E}_i, \Phi_{Ai},\bar{\Phi}_{Ai}, -\infty}   
\mathcal{D} T 
\mathcal{D} \bold{E} \mathcal{D}\bar{\tau} \mathcal{D}\Phi_A  \mathcal{D}\bar{\Phi}_A
\mathcal{D}c \mathcal{D}b
\int 
\mathcal{D} p_T
\\
&\qquad\times \exp \Biggl(- \int_{-\infty}^{\infty} dt \Bigl(-i p_{T}(t) \frac{d}{dt} T(t) 
 +\frac{d b(t)}{dt} \frac{d (T(t) c(t))}{dt}
+\lambda\bar{\rho}\frac{1}{T(t)}(\frac{d \bar{\tau}(t)}{dt})^2
+\int d\bar{\sigma} d^4 \bar{\theta}\bar{\bold{E}} T(t) \Phi_A' \bar{\Phi}_A'  \Bigr) \Biggr). 
\ea
\ee
where we have redefined as $c(t) \to T(t) c(t)$. $\bold{Z}_0$ represents an overall constant factor, and we will rename it $\bold{Z}_1, \bold{Z}_2, \cdots$ when the factor changes in the following.
This path integral is obtained if 
\begin{equation}
F_1(t):=\frac{d}{dt}T(t)=0 \label{superF1gauge}
\end{equation}
 gauge is chosen in 
\be
\ba
\Delta_F&(\Phi_{Af},\bar{\Phi}_{Af};  \Phi_{Ai},\bar{\Phi}_{Ai} | \bold{E}_f, ; \bold{E}_i) 
\\
&=
\bold{Z}_1\int^{\bold{E}_f, \Phi_{Af},\bar{\Phi}_{Af}, \infty }_{\bold{E}_i, \Phi_{Ai},\bar{\Phi}_{Ai}, -\infty}   
\mathcal{D} T 
\mathcal{D} \bold{E} \mathcal{D}\bar{\tau}\mathcal{D}\Phi_A  \mathcal{D}\bar{\Phi}_A
\int 
\exp \Biggl(- \int_{-\infty}^{\infty} dt \Bigl(
+\lambda\bar{\rho}\frac{1}{T(t)}(\frac{d \bar{\tau}(t)}{dt})^2
+\int d\bar{\sigma} d^4 \bar{\theta}\bar{\bold{E}} T(t) \Phi_A' \bar{\Phi}_A'  \Bigr) \Biggr), \label{manifest1dimdiff}
\ea
\ee
which has a manifest one-dimensional diffeomorphism symmetry with respect to $t$, where $T(t)$ is transformed as an einbein \cite{PhysRevD.44.3230}.


Under $\frac{d \bar{\tau}}{d \bar{\tau}'}=T(t)$, $T(t)$ disappears  in (\ref{manifest1dimdiff}) as in \cite{Sato:2017qhj}, and we obtain 
\be
\ba
\Delta_F &(\Phi_{Af},\bar{\Phi}_{Af};  \Phi_{Ai},\bar{\Phi}_{Ai} | \bold{E}_f, ; \bold{E}_i) 
\\
&=
\bold{Z}_2\int^{\bold{E}_f, \Phi_{Af},\bar{\Phi}_{Af}, \infty }_{\bold{E}_i, \Phi_{Ai},\bar{\Phi}_{Ai}, -\infty}   
\mathcal{D} \bold{E} \mathcal{D}\bar{\tau} \mathcal{D}\Phi_A  \mathcal{D}\bar{\Phi}_A
\int 
\exp \Biggl(- \int_{-\infty}^{\infty} dt \Bigl(
+\lambda\bar{\rho}(\frac{d \bar{\tau}(t)}{dt})^2
+\int d\bar{\sigma} d^4 \bar{\theta}\bar{\bold{E}}  \Phi_A'' \bar{\Phi}_A''  \Bigr) \Biggr),
\label{superpathint2}
\ea
\ee
where $\Phi_A''$ and $\bar{\Phi}_A''$ are given by replacing $\frac{\partial}{\partial \bar{\tau}}$ with $\frac{\partial}{\partial t}$ in $\Phi_A$ and $\bar{\Phi}_A$, respectively. This action is still invariant under the diffeomorphism with respect to t if $\bar{\tau}$ transforms in the same way as $t$. 

\begin{figure}[htb]
\centering
\includegraphics[width=6cm]{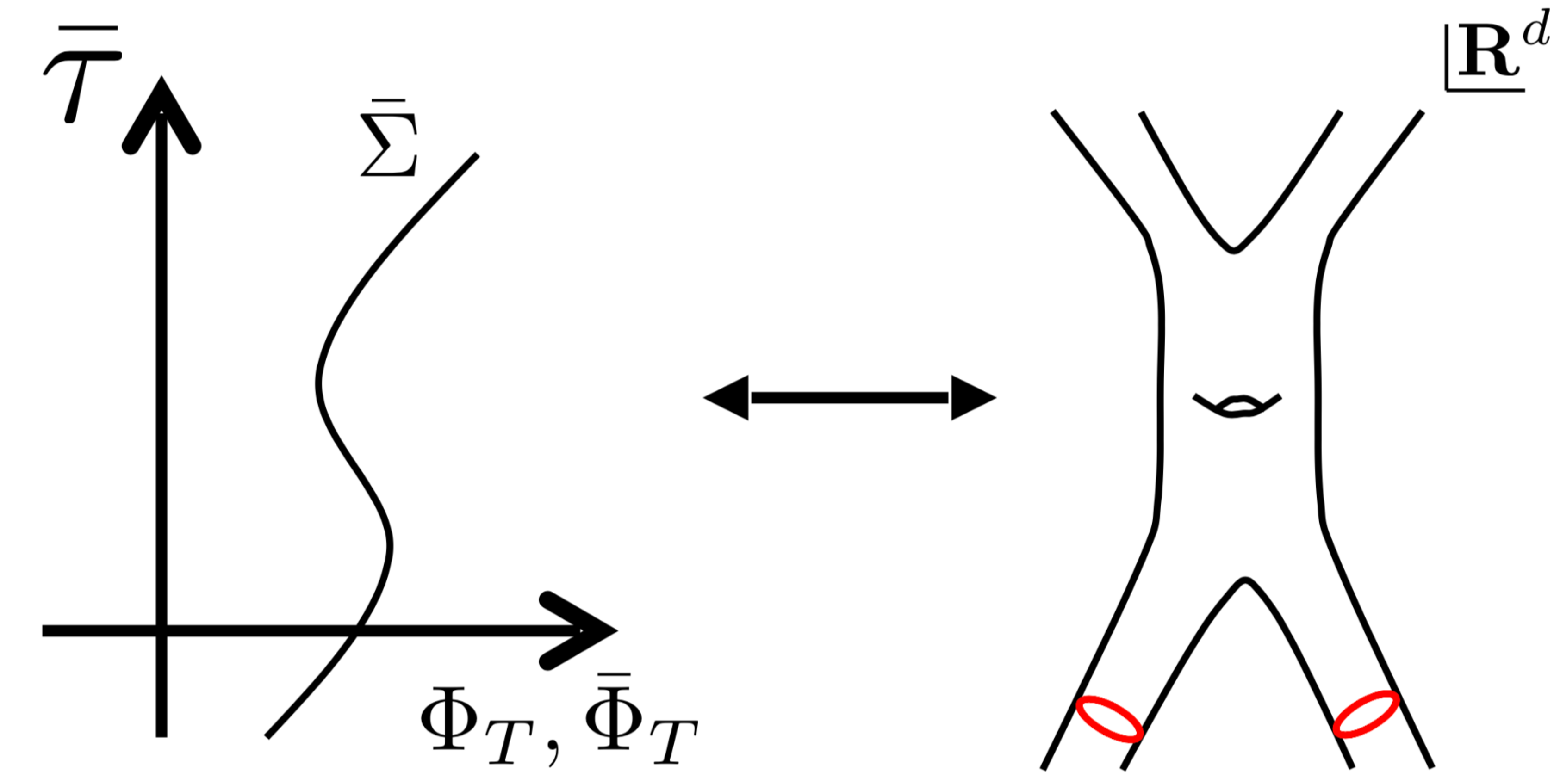}
\caption{A path and a topological super Riemann surface. The line on the left is a trajectory in the path integral. The trajectory parametrized by $\bar{\tau}$ from $\bar{\tau}=-\infty$ to $\bar{\tau}=\infty$, represents a topological super Riemann surface with fixed punctures in $\bold{R}^{d}$ on the right.} 
\label{Pathintegral}
\end{figure}

If we choose a different gauge
\begin{equation}
F_2(t):=\bar{\tau}-t=0, \label{superF2gauge}
\end{equation} 
in (\ref{superpathint2}), we obtain 
\be
\ba
\Delta_F &(\Phi_{Af},\bar{\Phi}_{Af};  \Phi_{Ai},\bar{\Phi}_{Ai} | \bold{E}_f, ; \bold{E}_i) 
 \\
&=
\bold{Z}_3\int^{\bold{E}_f, \Phi_{Af},\bar{\Phi}_{Af}, \infty }_{\bold{E}_i, \Phi_{Ai},\bar{\Phi}_{Ai}, -\infty}   
\mathcal{D} \bold{E} \mathcal{D}\bar{\tau} \mathcal{D}\Phi_A  \mathcal{D}\bar{\Phi}_A
\int 
\mathcal{D} \alpha \mathcal{D}c  \mathcal{D}b 
 \\
&\qquad \times
\exp \Biggl(- \int_{-\infty}^{\infty} dt \Bigl(\alpha(t) (\bar{\tau}-t) +b(t)c(t)(1-\frac{d \bar{\tau}(t)}{dt}) +\lambda \bar{\rho}(\frac{d \bar{\tau}(t)}{dt})^2 
+\int d\bar{\sigma} d^4 \bar{\theta}\bar{\bold{E}} \Phi_A'' \bar{\Phi}_A''   \Bigr) \Biggr) \\
&=
\bold{Z}
\int^{\bold{E}_f, \Phi_{Af},\bar{\Phi}_{Af}}_{\bold{E}_i, \Phi_{Ai},\bar{\Phi}_{Ai}} 
\mathcal{D} \bold{E}  \mathcal{D}\Phi_A  \mathcal{D}\bar{\Phi}_A
\exp \Biggl(- \int_{-\infty}^{\infty} d\bar{\tau} 
\Bigl(
\frac{1}{4 \pi}\int d\bar{\sigma} \sqrt{\bar{h}}
\lambda \bar{R}
+\int d\bar{\sigma} d^4 \bar{\theta} \bar{\bold{E}} \Phi_A \bar{\Phi}_A  \Bigr) \Biggr). \\
\ea
\ee
In the second equality, we have redefined as $c(t)(1-\frac{d \bar{\tau}(t)}{dt}) \to c(t)$ and integrated out the ghosts. The path integral is defined over all possible two-dimensional topological super Riemannian manifolds with fixed punctures in $\bold{R}^{d}$ as in Fig. \ref{Pathintegral}. 
By using the two-dimensional super diffeomorphism and super Weyl invariance of the action, we obtain
\begin{equation}
\Delta_F(\Phi_{Af},\bar{\Phi}_{Af};  \Phi_{Ai},\bar{\Phi}_{Ai} | \bold{E}_f, ; \bold{E}_i) 
=
\bold{Z}
\int^{\bold{E}_f, \Phi_{Af},\bar{\Phi}_{Af}}_{\bold{E}_i, \Phi_{Ai},\bar{\Phi}_{Ai}} 
\mathcal{D} \bold{E} \mathcal{D}\Phi_A  \mathcal{D}\bar{\Phi}_A
e^{-\lambda \chi} e^{-\int d^2\sigma d^4 \bar{\theta} \bar{\bold{E}} \Phi_A \bar{\Phi}_A },
\label{cSuperLast}
\end{equation} 
where $\chi$ is the Euler number of the two-dimensional Riemannian manifold. This is the all order perturbative partition function  of the A model for topological strings in  the flat target space itself \cite{Witten, Dijkgraaf:1990qw, Bershadsky:1993cx, Hori:1994nb}.

\section{Conclusion and discussion}

In this paper, we first defined topological string geometry theory by twisting string geometry theory.  From the single theory, we derived both the A and B models of the topological strings in  the flat target space, by considering fluctuations around the type A and B string manifolds covered by the A and B charts, respectively. This fact implies that we see the mirror symmetry in topological string geometry theory perturbatively.

The string coupling constant $g$ is not a parameter of non-perturbatively formulated string theory, but an expectation value of the dilaton. Actually, $g$ is not a parameter of the string geometry theory, too. $g=e^{\lambda}$ is a free parameter of the background solution (\ref{solution}) in string geometry theory when we derive the partition function of the perturbative topological string. During the derivation, we take a limit of another free parameter of the solution: $\alpha \to 0$. Therefore, it is natural to identify $\alpha$ as $e^{-\frac{1}{g}}$. We expect to obtain right non-perturbative corrections to the partition function by evaluating the contribution of the $\alpha$ expansions to the propagator in the string geometry theory, corresponding to the perturbative partition function.

Concretely, we can calculate the non-perturbative corrections \cite{progress}, as follows. In \cite{Sato:2020szq},  perturbative string theories are derived in Newtonian limits of string geometry theory including arbitrary fields. The post Newtonian expansion is the $\alpha$ expansion and can be identified with the $e^{-\frac{1}{g}}$ expansion.  Thus, by moving to the first quantization formalism in topological string geometry theory as in this paper,  we obtain $e^{-\frac{n}{g}}$ corrections to the perturbative $g^m$ contributions of the partition function. This behaviour is consistent with the results of the resurgence on the asymptotic expansion in the string coupling of the partition function in the topological string theory. That is, the summation over $m$ and $n$ of the $e^{-\frac{n}{g}}$ corrections to the $g^m$ contributions of the partition function equals to the summation over  $n$ and $m$ of the perturbative $g^m$ expansion in the $n$-th instanton background, whose factor is given by $e^{-\frac{n}{g}}$.  We can calculate these corrections by using the localization in the first quantization formalism of topological string geometry theory,  and compare them with the non-perturbative partition function which are conjectured by using dualities in \cite{Lockhart:2012vp, Hatsuda:2013oxa} and are coincident with the results of the resurgence in \cite{Pasquetti:2009jg, Hatsuda:2015oaa} . We also calculate them in a geometric approach as follows. We obtain corrections to the condition that curves are holomorphic as a result of the above localization.  Then, we obtain corrections to the definition of the Gromov-Witten invariants and calculate corrections to Gromov-Witten potential, which corresponds to the partition function at the zero-th order in the genus expansion. We compare them with the result in Physics. In the case that the target space is $C^3/Z_2$, we can calculate the corrections because  $C^3/Z_2$ can be treated as an orbifold of $C^3$ in topological string geometry theory\footnote{We can derive the perturbative topological string theory on $C_3/Z_2$ by patching open sets of the subspace whose target space is $C_3/Z_2$ in the model space and considering the fluctuations around the $Z_2$ orbifold of the perturbative vacuum solution (\ref{solution}).}, whereas it can be treated as a limit of a blow-up geometry in the conjecture, in the result of the resurgence, and in symplectic geometry.

Here, we discuss how to derive perturbative string theories on more general backgrounds. First, the perturbative vacuum solution (\ref{solution}) is a solution even if $v(\Phi_A(\bar{\tau}), \bar{\Phi}_A(\bar{\tau}))$ is generalized to arbitrary holomorphic plus anti-holomorphic functions. It is an immediate task to clarify on which K\"ahler manifolds topological string theories are reproduced from the fluctuations around the generalized backgrounds. Second, M. Honda and M. S. found that general configurations of the fields of a supergravity are included in configurations of the fields of the string geometry in \cite{preparation}. Similarly, we expect that general K\"ahler manifolds are included in configurations of the fields of the topological string geometry and that we can reproduce topological string theories on various compact and non-compact Calabi-Yau manifolds. 

Because string backgrounds are included in configurations of the fields of the string geometry \cite{preparation}, we expect that instantons of the string geometry reduce to instantons of the string backgrounds and instanton effects of string geometry give non-perturbative effects in string theory where a string background changes to another.

\section*{Acknowledgement}
We would like to thank
K. Hashimoto, S. Iso, H. Itoyama, H. Kawai, T. Kugo, T. Misumi, J. Nishimura, K. Ohta, N. Sakai, A. Tsuchiya, T. Yoneya, 
and especially 
S. Yamaguchi
for long and valuable discussions. 
 Y.S. thanks Interdisciplinary Center for Theoretical Study, University of Science and Technology of China for hospitality during his visit. The work of Y.S. is supported in part by the JSPS Research Fellowship for Young Scientists (No. JP17J00828).

\appendix

\section{Superfield formalism in topological string theory} \label{TS}
String geometry is described in terms of superfields. Then, we will introduce a superfield formalism in the topological string theory in this appendix. 

Two-dimensional $\mathcal{N}=(2,2)$ chiral matter coupled to supergravity in superspace is formulated in \cite{Grisaru:1995dr, Gates:1995du} , where the action is given by 
\be
\ba
S_{\text{sugra}} 
&=
\int {\rm d}^2 \sigma d^4 \theta \bold{E}' \eta_{I\bar{J}}\Phi^{' I} \bar{\Phi}^{' \bar{J}}
\\
&
=\int {\rm d}^2 z 
\eta_{I\bar{J}}
(\partial_{z^*} \phi^I \partial_z \bar{\phi}^{\bar{J}}
+F^I  \bar{F}^{\bar{J}} 
+
 {\rm i}
\psi_+^I D_z \psi_{\dot{+}}^{\bar{J}} 
 +
 {\rm i}
\psi_-^I D_{z^*} \psi_{\dot{-}}^{\bar{J}}
 \\
 &\qquad
+\partial_{z^*} \phi^I \psi_{z}^{\dot{+}} \psi_{\dot{+}}^{\bar{J}} 
+\partial_{z} \phi^I \psi_{z^*}^{\dot{-}} \psi_{\dot{-}}^{\bar{J}} 
+\partial_{z^*} \bar{\phi}^{\bar{J}}  \psi_{z}^{+} \psi_{+}^{I} 
+\partial_{z} \bar{\phi}^{\bar{J}}  \psi_{z^*}^{-} \psi_{-}^{I} 
\\
&\qquad
+\frac{1}{2}
(\psi^+_{z^*} \psi_+ +\psi^-_{z^*} \psi_-) \psi_z^{\dot{+}} \psi_{\dot{+}}
+\frac{1}{2}
(\psi^+_{z} \psi_+ +\psi^-_{z} \psi_-) \psi_{z^*}^{\dot{-}} \psi_{\dot{-}}
\\
&\qquad
+\frac{1}{2}
(\psi^{\dot{+}}_{z^*} \psi_{\dot{+}} +\psi^{\dot{-}}_{z^*} \psi_{\dot{-}}) \psi_z^+ \psi_+
+\frac{1}{2}
(\psi^{\dot{+}}_{z} \psi_{\dot{+}} +\psi^{\dot{-}}_{z} \psi_{\dot{-}}) \psi_{z^*}^- \psi_-).
\ea
\ee
To this system, we perform formal topological A and B twists where  the half-integer spins are changed to the integer spins as in the same changes that are results of the topological twists of the sigma models. Explicitly, the super fields are changed as  
\begin{equation}
\bold{E}' \to \bold{E}, \, \Phi^{' I} \to \Phi_T^I, \, \bar{\Phi}^{' \bar{I}} \to \bar{\Phi}_T^{\bar{I}} \; (T=A,B), \label{Phi}
\end{equation}
where  the coordinates are 
\be
\ba
&\theta^+ \to \theta^{z^*}, ~ \theta^- \to \theta , ~ \theta^{\dot{+}} \to \bar{\theta}, ~ \theta^{\dot{-}} \to \theta^z,
\label{mod_supfiA}
\ea
\ee
for $T=A$ twist and
\be
\ba
&\theta^+ \to \theta^{z^*}, ~ \theta^- \to \theta^z , ~ \theta^{\dot{+}} \to \bar{\theta}, ~ \theta^{\dot{-}} \to \theta,
\label{mod_supfiB}
\ea
\ee
for $T=B$ twist, and the component fields are
\be
\ba
&\psi^I_+ \to  \rho_{z^*}^{I}  ,~ \psi^I_-  \to  \chi^I ,
\\
&\psi^{\bar{I}}_{\dot{+}}  \to   \bar{\chi}^{\bar{I}},~ \psi^{\bar{I}}_{\dot{-}} \to  \bar{\rho}_z^{\bar{I}},
\\
&\psi_{z}^{\dot{-}} \to 0, ~ \psi_{z^*}^{\dot{-}} \to -i \chi^z_{z^*}, ~\psi_z^+ \to -i \chi_z^{z^*}, ~ \psi_{z^*}^+ \to 0
\\
&\psi_{z}^{\dot{+}} \to 0, ~ \psi_{z^*}^{\dot{+}} \to 0, ~ \psi_{z}^- \to 0, ~ \psi_{z^*}^- \to 0,
\ea
\ee
for $T=A$ and 
\be
\ba
&\psi^I_+ \to  \rho_{z^*}^{I}  ,~ \psi^I_-  \to   \bar{\rho}_z^I,
\\
&\psi^{\bar{I}}_{\dot{+}}  \to   \bar{\chi}^{\bar{I}},~ \psi^{\bar{I}}_{\dot{-}} \to \chi^{\bar{I}},
\\
&\psi_{z}^{\dot{-}} \to 0, ~ \psi_{z^*}^{\dot{-}} \to 0, ~\psi_z^+ \to -i \chi_z^{z^*}, ~ \psi_{z^*}^+ \to 0
\\
&\psi_{z}^{\dot{+}} \to 0, ~ \psi_{z^*}^{\dot{+}} \to 0, ~ \psi_{z}^- \to 0, ~ \psi_{z^*}^- \to  -i \chi^z_{z^*},
\ea
\ee
for $T=B$ where some component fields  are set to zero. As a result, $\chi$ and $ \bar{\chi}$ transform as scalar fields  whereas $ \rho_{z^*}$ and $\bar{\rho}_z$ transform as chiral vector fields, and $\chi_z^{z^*}$ and $\chi^z_{z^*}$ transform as spin 2 gravitinos.

The new superfields $\bold{E}$, $\Phi_T$ and $\bar{\Phi}_T$ satisfy
\be
\ba
S_{A} 
&=
\int {\rm d}^2 \sigma d^4 \theta \bold{E} \eta_{I\bar{J}}\Phi_A^I \bar{\Phi}_A^{\bar{J}}
\\
&=
\int {\rm d}^2 z 
\eta_{I\bar{J}}
(\partial_{z^*} \phi^I \partial_z \bar{\phi}^{\bar{J}}
+F^I  \bar{F}^{\bar{J}}
 +
 {\rm i}
\rho_{z^*}^{I} D_z \bar{\chi}^{\bar{J}}
 +
 {\rm i}
\bar{\rho}_z^{\bar{J}} D_{z^*}  \chi^I   
 \\
 &\qquad
 +
 {\rm i}
\rho_{z^*}^{I} \chi_z^{z^*} \partial_{z^*} \bar{\phi}^{\bar{J}} 
 +
 {\rm i}
\bar{\rho}_z^{\bar{J}} \chi^z_{z^*} \partial_{z} \phi^I 
-\chi^z_{z^*}\chi_z^{z^*} \rho_{z^*}^{I} \bar{\rho}_z^{\bar{J}}
),
\label{top_actionA}
\ea
\ee
\be
\ba
S_{B} 
&=
\int {\rm d}^2 \sigma d^4 \theta \bold{E} \eta_{I\bar{J}}\Phi_B^I \bar{\Phi}_B^{\bar{J}}
\\
&=
\int {\rm d}^2 z 
\eta_{I\bar{J}}
(\partial_{z^*} \phi^I \partial_z \bar{\phi}^{\bar{J}}
+F^I  \bar{F}^{\bar{J}}
 +
 {\rm i}
\rho_{z^*}^{I} D_z \bar{\chi}^{\bar{J}}
 +
 {\rm i}
\bar{\rho}_z^I D_{z^*}  \chi^{\bar{J}}
 \\
 &\qquad
 +
 {\rm i}
\rho_{z^*}^{I} \chi_z^{z^*} \partial_{z^*} \bar{\phi}^{\bar{J}} 
 +
 {\rm i}
\bar{\rho}_z^I \chi^z_{z^*} \partial_{z} \bar{\phi}^{\bar{J}} 
),
\label{top_actionB}
\ea
\ee
which are the actions of the A and B models of topological strings\cite{Dijkgraaf:1990qw, Witten, Bershadsky:1993cx, Hori:1994nb}, respectively. Therefore, the new superfields $\bold{E}$, $\Phi_T^I$ and $\bar{\Phi}_T^{\bar{I}}$ define superfield formalisms of the topological string theories.

\providecommand{\href}[2]{#2}\begingroup\raggedright\endgroup

\end{document}